 \documentclass[proof]{pasj01}
\Received{$\langle$reception date$\rangle$}
\Accepted{$\langle$acception date$\rangle$}
\Published{$\langle$publication date$\rangle$}
\usepackage{lineno}

\begin{document}

\newcommand{\ncconh}{\mbox{NCCONH$_2$}}

\title{Tentative detection of cyanoformamide NCCONH$_2$ in space}

\author{Juan \textsc{Li} \altaffilmark{1,2} }
\email{lijuan@shao.ac.cn}
\author{ Donghui  \textsc{Quan} \altaffilmark{3,4}}
\author{  Junzhi  \textsc{Wang} \altaffilmark {5}}
\author{  Xia  \textsc{Zhang} \altaffilmark{6}}
\author{ Xing  \textsc{Lu} \altaffilmark {1,2} }
\author{  Qian  \textsc{Gou} \altaffilmark {7} }
\author{  Feng  \textsc{Gao} \altaffilmark {8} }
 \author{ Yajun  \textsc{Wu} \altaffilmark {1,2}  }
\author{  Edwin  \textsc{Bergin} \altaffilmark{9}  }
\author{  Shanghuo  \textsc{Li} \altaffilmark{10} }
\author{  Zhiqiang  \textsc{Shen} \altaffilmark{1,2}  }
\author{  Fujun  \textsc{Du} \altaffilmark{2,11}  }
\author{ Meng  \textsc{Li} \altaffilmark{7}   }
\author{ Siqi  \textsc{Zheng} \altaffilmark{1,2,12}   }
\author{ Xingwu  \textsc{Zheng} \altaffilmark{13} }

\altaffiltext{1}{Department of Radio Science and Technology, Shanghai Astronomical observatory, 80 Nandan RD, Shanghai 200030, China}
\altaffiltext{2}{Key Laboratory of Radio Astronomy, Chinese Academy of Sciences, China}
\altaffiltext{3}{Research Center for Intelligent Computing Platforms, Zhejiang Laboratory, Hangzhou 311100, PR China}
\altaffiltext{4}{Department of Chemistry, Eastern Kentucky University, Richmond, KY 40475, USA}
\altaffiltext{5}{Guangxi key Laboratory for Relativistic Astrophysics, Department of Physics, Guangxi University, Nanning 530004, PR China}
\altaffiltext{6}{Xinjiang Astronomical Observatory, Chinese Academy of Sciences, 150 Science 1-Street, Urumqi 830011, PR China }
\altaffiltext{7}{School of Chemistry and Chemical Engineering, Chongqing University, Daxuecheng South Rd. 55, 401331, Chongqing, PR China}
\altaffiltext{8}{Hamburger Sternwarte, Universität Hamburg, Gojenbergsweg 112, 21029, Hamburg, Germany }
\altaffiltext{9}{Department of Astronomy, University of Michigan, Ann Arbor, MI 48109, USA}
\altaffiltext{10}{Max Planck Institute for Astronomy, Königstuhl 17, D-69117 Heidelberg, Germany}
\altaffiltext{11}{Purple Mountain Observatory, Chinese Academy of Sciences, Nanjing 210034, PR China}
\altaffiltext{12}{University of Chinese Academy of Sciences, 19A Yuquanlu, Beijing 100049, PR China}
\altaffiltext{13}{School of Astronomy and Space Science, Nanjing University, Nanjing 210093, China}

\KeyWords{ISM: abundances ${}_1$ ---  ISM: clouds${}_2$ ---SM: individual (Sagittarius B2)${}_3$ --- ISM: 
molecules${}_4$ --- radio lines: ISM${}_5$  }

\maketitle

\begin{abstract}
The peptide-like molecules, cyanoformamide (\ncconh), is the cyano (CN) derivative of formamide (NH$_2$CHO). It is known to play a role in the synthesis of nucleic acid precursors under prebiotic conditions. In this paper, we present a tentative detection of \ncconh\ in the interstellar medium (ISM) with the Atacama Large Millimeter/submillimeter Array (ALMA) archive data. Ten unblended lines of \ncconh\ were seen around 3$\sigma$ noise levels toward Sagittarius B2(N1E), a position that is slightly offset from the continuum peak. The column density of \ncconh\ was estimated to be 2.4$\times 10^{15}$cm$^{-2}$, and the fractional abundance of \ncconh\ toward Sgr B2(N1E) was $6.9\times10^{-10}$. The abundance ratio between NCCONH$_2$ and NH$_2$CHO is estimated to be $\sim$0.01. We also searched for other peptide-like molecules toward Sgr B2(N1E). The abundances of NH$_2$CHO, CH$_3$NCO and CH$_3$NHCHO toward Sgr B2(N1E) were about one tenth of those toward Sgr B2(N1S), while the abundances of CH$_3$CONH$_2$ was only one twentieth of that toward Sgr B2(N1S). 
\end{abstract}

\section{Introduction}

Peptide bonds, -NHCO-, are bridges that connect amino acids to form proteins, which are the basis of life on Earth \citep{Kaiser13}. Studies of molecules with peptide-like bonds are thus important for our understanding of the origin of life \citep{Halfen11, Belloche17}. A number of peptide-like molecules have been detected in space. Formamide (NH$_2$CHO), which is a potential precursor of various organic compounds essential to life \citep{Lopez2019}, was the first peptide-like molecule detected in space. \citet{Rubin71} reported the detection of formamide in Sagittarius B2 (hereafter Sgr B2) with the 140-foot telescope of the National Radio Astronomy Observatory (NRAO). After that, formamide was detected in a number of massive star-forming regions, including Orion KL \citep{Turner1989, Turner1991}, W3(H$_2$O) \citep{Bisschop2007}, and W51e1/e2 \citep{Suzuki2018} and so on. It was also found in some solar-mass protostellar cores, including the well-studied protobinary IRAS 16293-1622 \citep{Coutens16}, the protostellar shock region L1157-B1 \citep{Mendoza14} and so on. Formamide was also detected in the gas and ices of some comets \citep{Altwegg17}. The next peptide-like molecule, acetamide (CH$_3$CONH$_2$), was also first detected toward Sgr B2(N) \citep{Hollis06, Halfen11}, and later detected in more high-mass star forming regions \citep{Ligterink20, Colzi21}. N-methylformamide (CH$_3$NHCHO), the isomer of acetamide, was detected towards Sgr B2(n1S), NGC 6334I,  G31.41+0.31 and so on \citep{Belloche17, Belloche19, Ligterink20, Colzi21, Zeng23}. Urea (NH$_2$C(O)NH$_2$), were first identified toward Sgr B2(N) with the ALMA \citep{Belloche17, Belloche19}, and later detected in molecular cloud G+0.693-0.027 in the Sgr B2 complex \citep{Jimenez20}. In addition, propionamide C$_2$H$_5$CONH$_2$ was tentatively detected toward Sgr B2(N1E) \citep{Li21}. These observations suggest that peptide-like molecules might be widespread in space. 


The organic species cyanoformamide (NCCONH$_2$) is the cyano (CN) derivative of formamide (NH$_2$CHO), a known interstellar molecule with a role in the synthesis of nucleic acid precursors under prebiotic conditions \citep{Winnewisser05}. Cyanoformamide could form from the reaction between CN and formamide \citep{Winnewisser05}. As both CN and formamide are observed to be abundant in the ISM, cyanoformamide is expected to be detectable in the ISM \citep{Winnewisser05}. At room temperature and normal terrestrial pressure, this molecule is rather unstable. The microwave and millimeter-wave spectrum of the gas-phase species was not studied until 2005 \citep{Christiansen05, Winnewisser05}. Recently, \citet{Colzi21} searched cyanoformamide toward hot core G31.41+0.31 but got a negative result. 

Sgr B2, the giant molecular cloud located in the Galactic central region, is the most massive star-forming region in the Galaxy. It has long been regarded as the best hunting grounds for complex organic molecules (COMs) due to the extraordinary molecular richness \citep{Belloche13, Li17, Li20}. Most O-bearing COMs were first detected toward Sgr B2 \citep{McGuire18, McGuire22}, such as the branched molecule i-C$_3$H$_7$CN \citep{Belloche14} and the chiral molecule CH$_3$CHCH$_2$O \citep{McGuire16}. There are two main sites of star formation regions in Sgr B2, namely Sgr~B2(N) and Sgr~B2(M), both of which host several dense, compact, hot cores that are rich in COMs \citep{Bonfand17}. Recently, \citet{Li21} found evidence for the possible presence of C$_2$H$_5$CONH$_2$, the largest peptide-like molecule, toward Sgr B2(N1E), suggesting that peptide-like molecules are abundant in this region.

In this paper, we present a tentative detection of NCCONH$_2$ toward Sgr B2(N1E). Section 2 introduces the observations and data reduction. Section 3 presents the observational results. Section 4 discusses the observing results and possible formation mechanisms of NCCONH$_2$. The summary of the work is presented in Section 5. 


\section{OBSERVATIONS AND DATA REDUCTION}\label{sec:obs}

The data used for this study were acquired from the ALMA Science archive of the ReMoCA survey (Re-exploring Molecular Complexity with ALMA) \citep{Belloche19}. The ReMoCA survey was conducted with ALMA during Cycle 4 between 2016 and 2017. Detailed description about the observations is presented in \citet{Belloche19}. This is a compete spectral line survey toward Sgr B2(N) covering from 84.1 to 114.4 GHz. Five spectral setups were used in total. The on-source time for each frequency windows range from 47 to 50 minutes. The spectral resolution is 0.488 MHz, which corresponds to a velocity resolution of 1.3-1.7 km s$^{-1}$ across the observing band. The phase center of the observations is $\alpha_{\rm J2000}=17^{\rm h}47^{\rm m}19.87^{\rm s}$, $\delta_{\rm J2000}=-28^\circ 22'16''$, which locates half way between the two main hot molecular hot cores Sgr B2(N1) and N2.

Our data reduction procedure has been introduced in \citet{Li21}. The data was calibrated using the standard ALMA data calibration pipeline with the Common Astronomy Software Applications package (CASA). CASA version 4.7.0-1 was used for the first spectral setup, while CASA version 4.7.2 was used for the other four spectral setups. The CASA version 5.6.1-8 was used to image the calibrated data. The quasar J1924-2914 was used to calibrate the bandpass for most of the data, while the quasar J1517-2422 was used to calibrate the bandpass for on execution in the second spectral setup. Quasars J1924-2914 or J1733-1304 were used to derive the absolute flux density scale. The quasar J1744-3116 was used to calibrate the phase and amplitude. The TCLEAN deconvolution algorithm in CASA was used to produce the images. Self-calibration is known to introduce artificial features into imaging by including manually chosen clean components if the structure of the target is complicated (i.e., if it is not a simple point source). Sgr B2(N) has complicated substructures \citep{Bonfand17, Belloche19}, therefore, self-calibration will likely introduce artifacts into images, which may affect spectral line identification, especially for the weak lines in our case. We did not perform self-calibration, but still achieved a sufficiently high imaging rms of 0.4 $\sim$ 1 mJy beam$^{-1}$, which is not much higher than that in \citet{Belloche19}. Therefore, we believe that this strategy is suitable for our purpose of identification of weak lines. 

Because of high number of spectral lines detected toward hot cores in Sgr B2, the determination of baseline is challenging. As pointed out by \citet{sanchez2018}, a broad frequency coverage is necessary to ensure the presence of enough line-free frequency intervals to determine the continuum level. They think that a bandwidth of at least 1 GHz is needed. \citet{sanchez2018} have simulated spectra that dominated by emission features, and eight methods were used to determine the continuum level of the spectra. We could see from Table 1 in \citet{sanchez2018} that all the method over-estimated the rms noise levels by 2\% to 113\%. We could see from Figure 1(b) in \citet{sanchez2018} that the continuum level of the spectra could be better determined with several groups of lowest values in the spectra after masking absorption lines. We first compared with spectra toward HII regions to mask absorption lines, then we chose several groups of lowest values to determine the continuum levels. This could properly determine the continuum levels for spectra dominated by emission lines.

The rms noise levels for the spectra window were determined using the median values of channel maps. We first investigated the rms noise levels in regions without either continuum emission or molecular lines. Then we investigated the rms noise level in regions with strong continuum emission, which are significantly larger than rms noise levels in regions without either molecular lines or continuum emission. The median values of rms noise levels were adopted in this paper.


\section{RESULTS}\label{sec:results}

\subsection{Identification of \ncconh}
\label{sec:results1}

Weeds in Gildas package was used for line identification and spectral modeling. The Jet Population Laboratory (JPL) \citep{Pickett98} and the Cologne Database for Molecular Spectroscopy (CDMS) \citep{CDMS2, Endres16} databases were used. The microwave and millimeter-wave spectrum of \ncconh\ was reported by \citet{Christiansen05} and  \citet{Winnewisser05}. The 3-mm emission of NCCONH$_2$ is modeled assuming LTE conditions with five parameters: column density, temperature, source size, velocity offset, and linewidth. Each spectral window of each observed setup was modeled separately to account for the varying angular resolution, but with a same set of parameters.

Approximately 120 NCCONH$_2$ lines were expected to be seen in the observed frequency ranges. Ten unblended transitions, and 21 partially blended transitions of NCCONH$_2$ were possibly seen toward Sgr B2(N1E) (Figure~\ref{f:sgrb2}), which is about 1.5$''$ to the east of the hot core Sgr~B2(N1). However, since their signal-to-noise levels are around 3$\sigma$, it is not possible to claim secure detection of \ncconh\ with these data. The large peptide-like molecule, propionamide has been tentatively found to be relatively abundant toward Sgr B2(N1E) \citep{Li21}. We also searched for NCCONH$_2$ toward other regions of Sgr~B2(N1). However, serious line-blending prevents detection of this molecule toward other directions. The tentatively detected lines are presented in Figures~\ref{f:clean} and ~\ref{f:partial}, and Table~\ref{tab:freq}. Though the  remaining $\sim$ 90 lines that match the considered frequency range and noise level threshold are contaminated by those of other species, the observed results do not contradict with the expected intensities. In Figure~\ref{f:clean}, the black dashed lines show the 3$\sigma$ noise levels adopted in Section 2. 

Because of low abundance of \ncconh\ and serious line blending in Sgr B2(N1), all of these unblended lines detected toward Sgr B2(N1E) suffer from line blending toward other directions. Thus we could not get the spatial distribution of \ncconh\ in Sgr B2(N1).

\subsection{Search of related molecules toward Sgr B2(N1E)}
\label{sec:results2}

Some related molecules were also searched toward Sgr B2(N1E). Based on observing results of \citet{Belloche19}, we searched for nine relatively clean and strong transitions of CH$_3$NCO v=0 toward Sgr B2(N1E). The CDMS database is used for the identification. Nine clean lines were detected. As nine lines are enough for the detection and modelling of molecule in Sgr B2, we did not search for all the CH$_3$NCO v=0 toward Sgr B2(N1E). The spectra were shown in Figure~\ref{f:ch3nco}. The intensity of CH$_3$NCO lines toward Sgr B2(N1E) were weaker than those toward Sgr B2(N1S) (See Figure A.6 in \citet{Belloche19}). 

Thirteen relatively clean and strong transitions of CH$_3$NHCHO v=0 were searched toward Sgr B2(N1E). The Lille Spectrascopic Database (https://lsd.univ-lille.fr/) is used for the identification. All of these lines were detected. The spectra were shown in Figure~\ref{f:ch3nhcho}. The intensity of CH$_3$NHCHO lines toward Sgr B2(N1E) were significantly weaker than those toward Sgr B2(N1S) (See Figure A.3 in \citet{Belloche19}). 

We also searched for urea (NH$_2$CONH$_2$) toward Sgr B2(N1E). To our knowledge, this molecule was only detected toward Sgr B2(N1S) \citep{Belloche19} and G+0.693-0.027 \citep{Jimenez20}, a molecular cloud located in the Sgr B2 complex. It was tentatively detected toward NGC6334I\citep{Ligterink20}. Based on observing results of \citet{Belloche19}, we searched for four clean and strong NH$_2$CONH$_2$ v=0 transitions toward Sgr B2(N1E). Unfortunately, all of these lines blend seriously with other molecules.

\subsection{Column densities of \ncconh\ and related molecules}

The column densities of \ncconh\ and related molecules are obtained by eye-fitting the spectra in Weeds. The physical size of the emission region is hard to determine, as the morphology of molecules in Sgr B2(N1) does not simply follow a 2D Gaussian profile \citep{Busch22}. The emitting size was assumed to be 2.3\arcsec, thus the beam filling factor is $\sim$1. With a linewidth measurement of 3.0 km s$^{-1}$, and an assumption of excitation temperature of 150 K (which is very close to what was obtained for molecules toward Sgr B2(N1S) by \citet{Belloche19}, the column density and centroid velocity are varied to fit the detected transitions. In this way, a column density of $2.4\times10^{15}$ cm$^{-2}$ was obtained for cyanoformamide (see Table~\ref{tab:column}). By adopting the H$_2$ column density of 3.5$\times 10^{24}$ cm$^{-2}$ derived with C$^{18}$O \citep{Li21}, the abundance relative to H$_2$ was estimated to be $6.9\times10^{-10}$ for NCCONH$_2$. 

We also modeled the emission of CH$_3$NCO and CH$_3$NHCHO toward Sgr B2(N1E) with the same size and excitation temperature (see Table~\ref{tab:column}). A column density of 2.4$\times 10^{16}$ cm$^{-2}$ was obtained for CH$_3$NCO. A column density of $2.1\times10^{16}$ cm$^{-2}$ was obtained for CH$_3$NHCHO.  

Based on the estimated column density of NCCONH$_2$ toward Sgr B2(N1E), the abundance ratio between NCCONH$_2$ and formamide is found to be $\sim$0.01. This value is consistent with result in \citet{Colzi21}. They found that the abundance ratio of NCCONH$_2$ to formamide was lower than 0.05 in G31.41+0.31.


\section{DISCUSSION}\label{sec:disc}

 \subsection{Comparison of Sgr B2(N1E) and Sgr B2(N1S)}
 
\begin{figure}
\centering
{\includegraphics[width=3.75in,angle=-90]{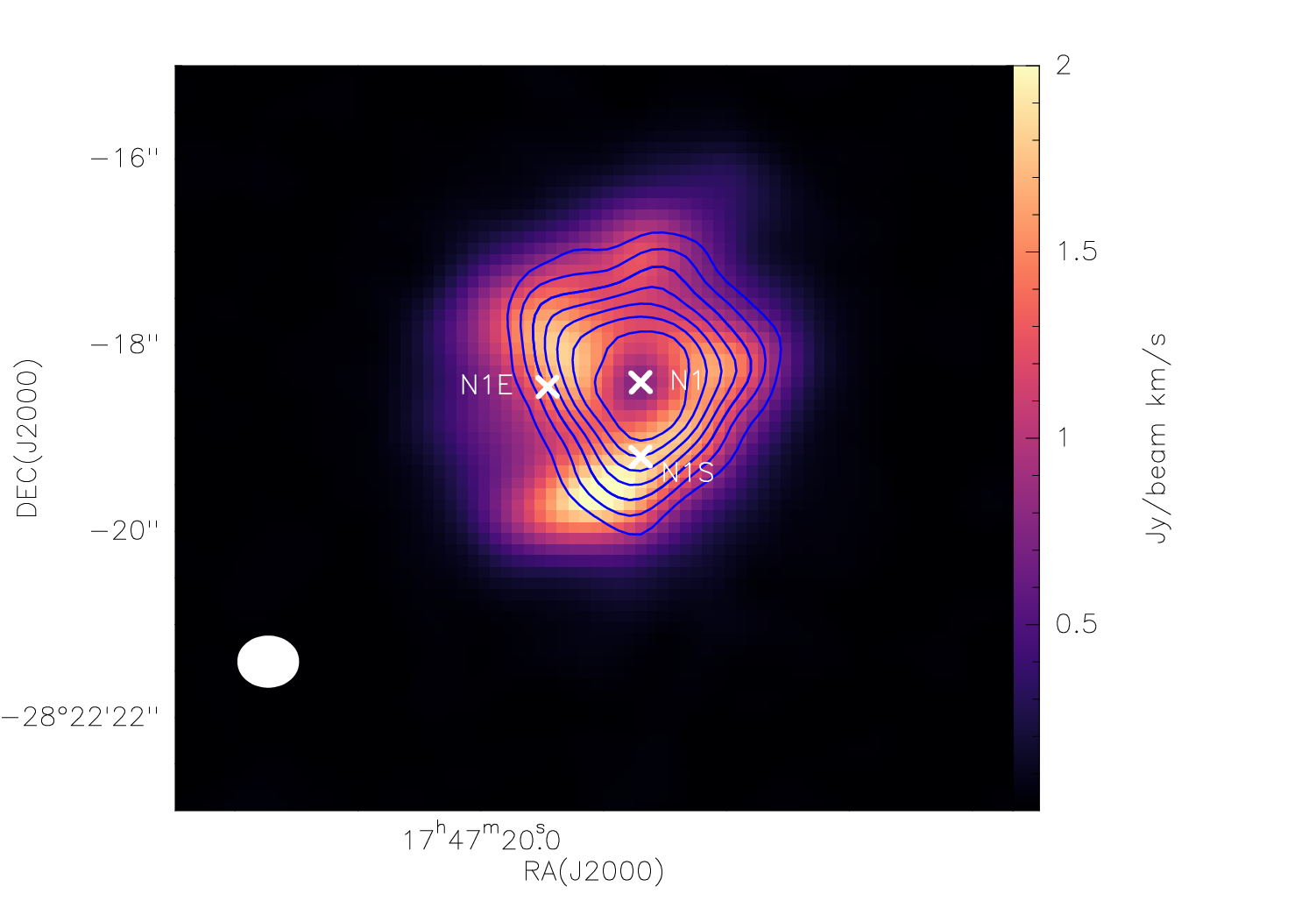}}
        \caption{Integrated intensity map of NH$_2$CHO and CH$_3^{13}$CN in Sgr~B2(N1). The integrated intensity map of NH$_2$CHO at 103.525 GHz are shown in color scale. The integrated intensity map of CH$_3^{13}$CN at 91.94 GHz is shown in contours, which represent 30\% to 90\% percent of peak integrated intensity. The white crosses indicate the position of Sgr B2(N1E), Sgr B2(N1S), and the center of Sgr B2(N1). The white ellipse shows the size of the respective synthetic beam. 
}
\label{f:sgrb2}
 \end{figure}

\begin{figure}
\centering
{\includegraphics[width=2.5in]{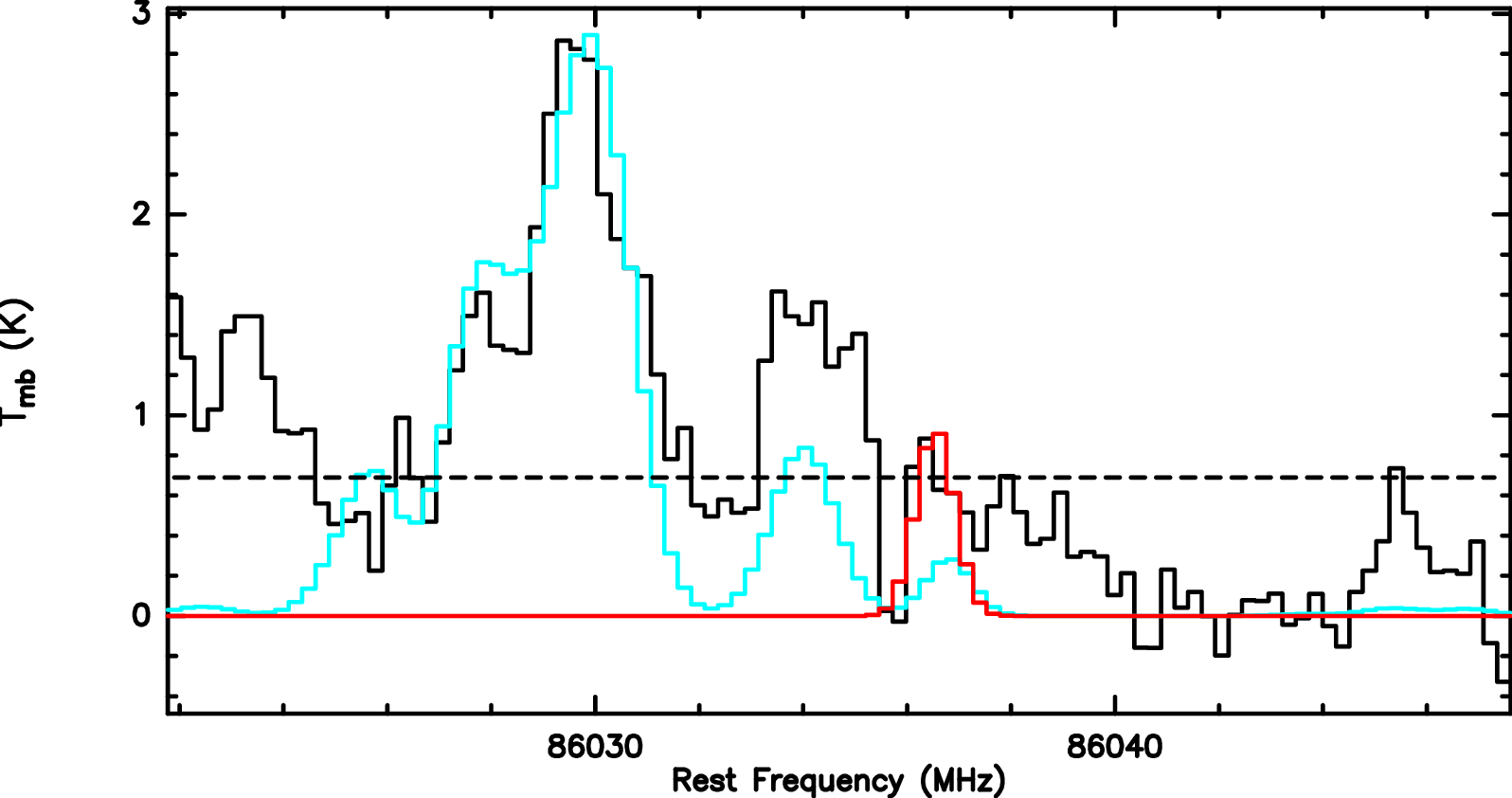}
 \includegraphics[width=2.5in]{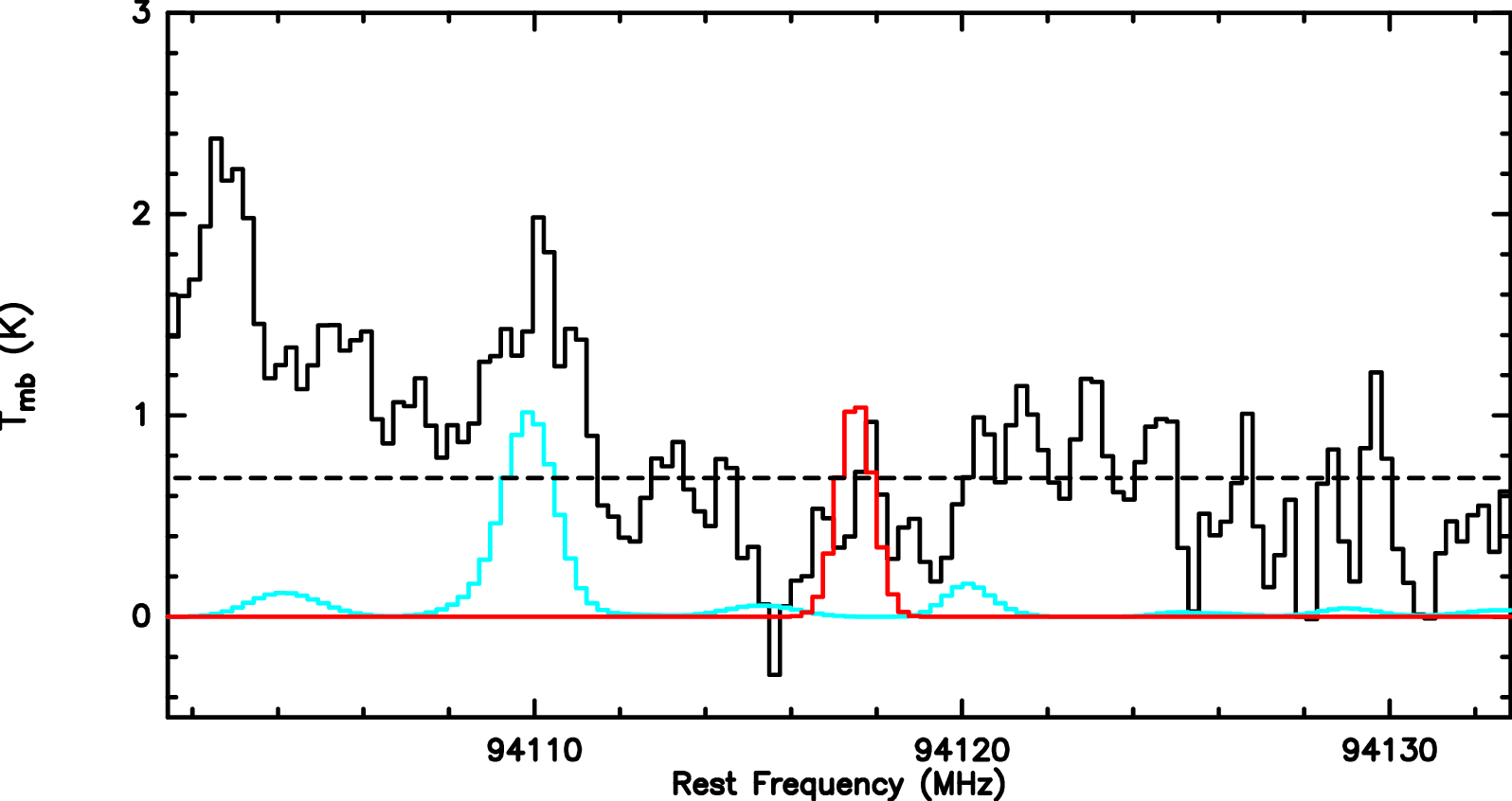}
\includegraphics[width=2.5in]{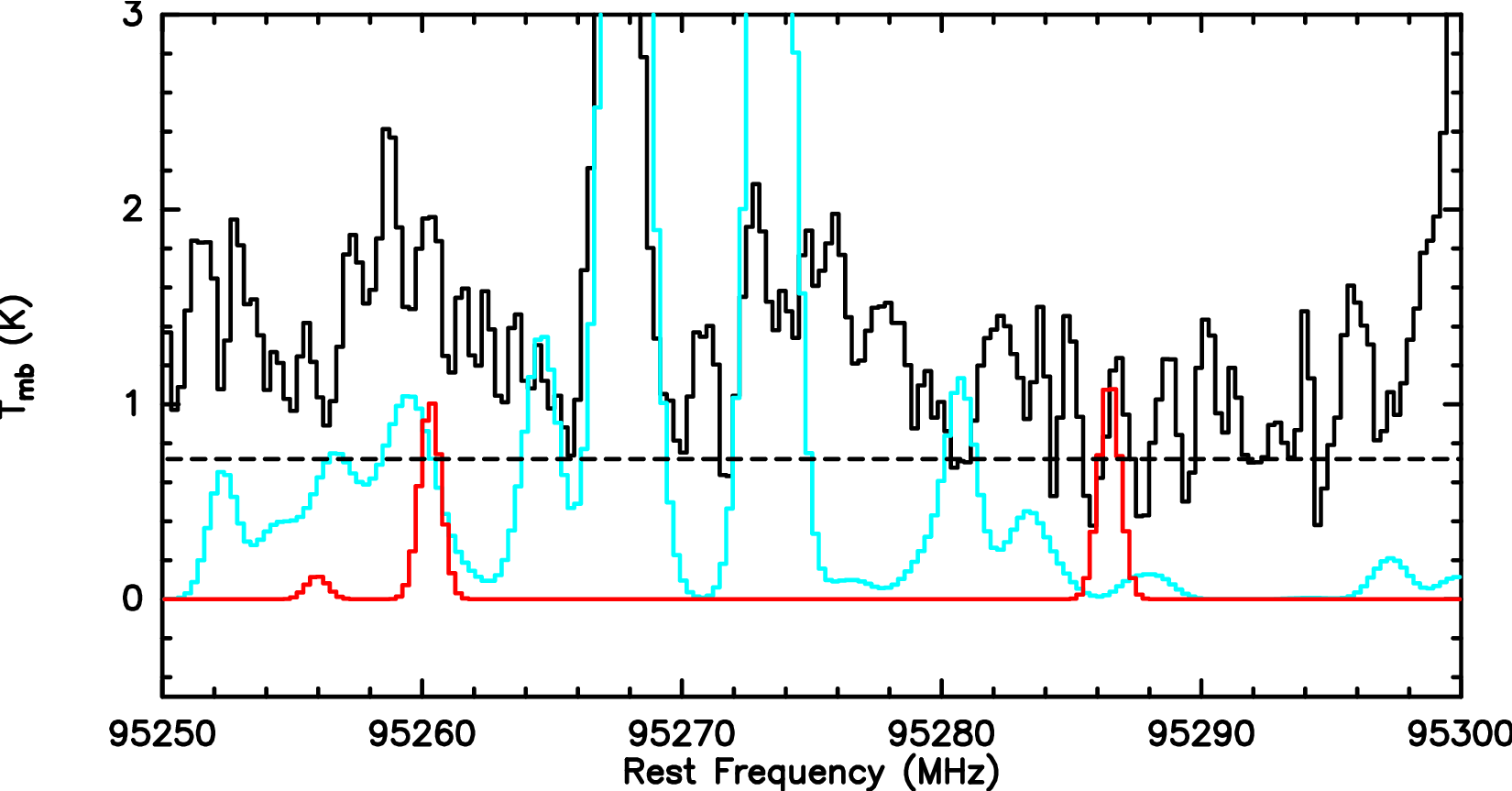}
 \includegraphics[width=2.5in]{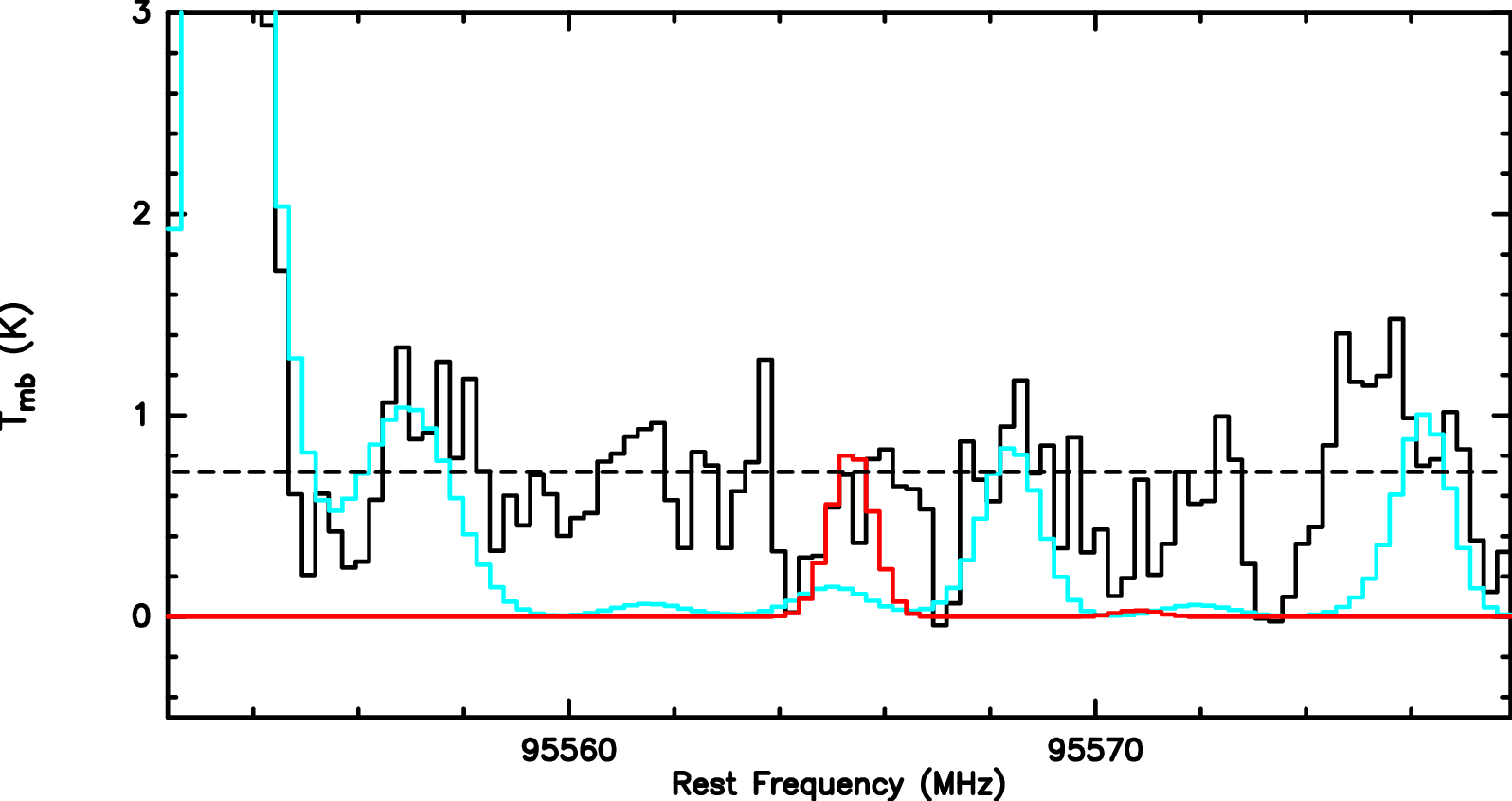} 
\includegraphics[width=2.5in]{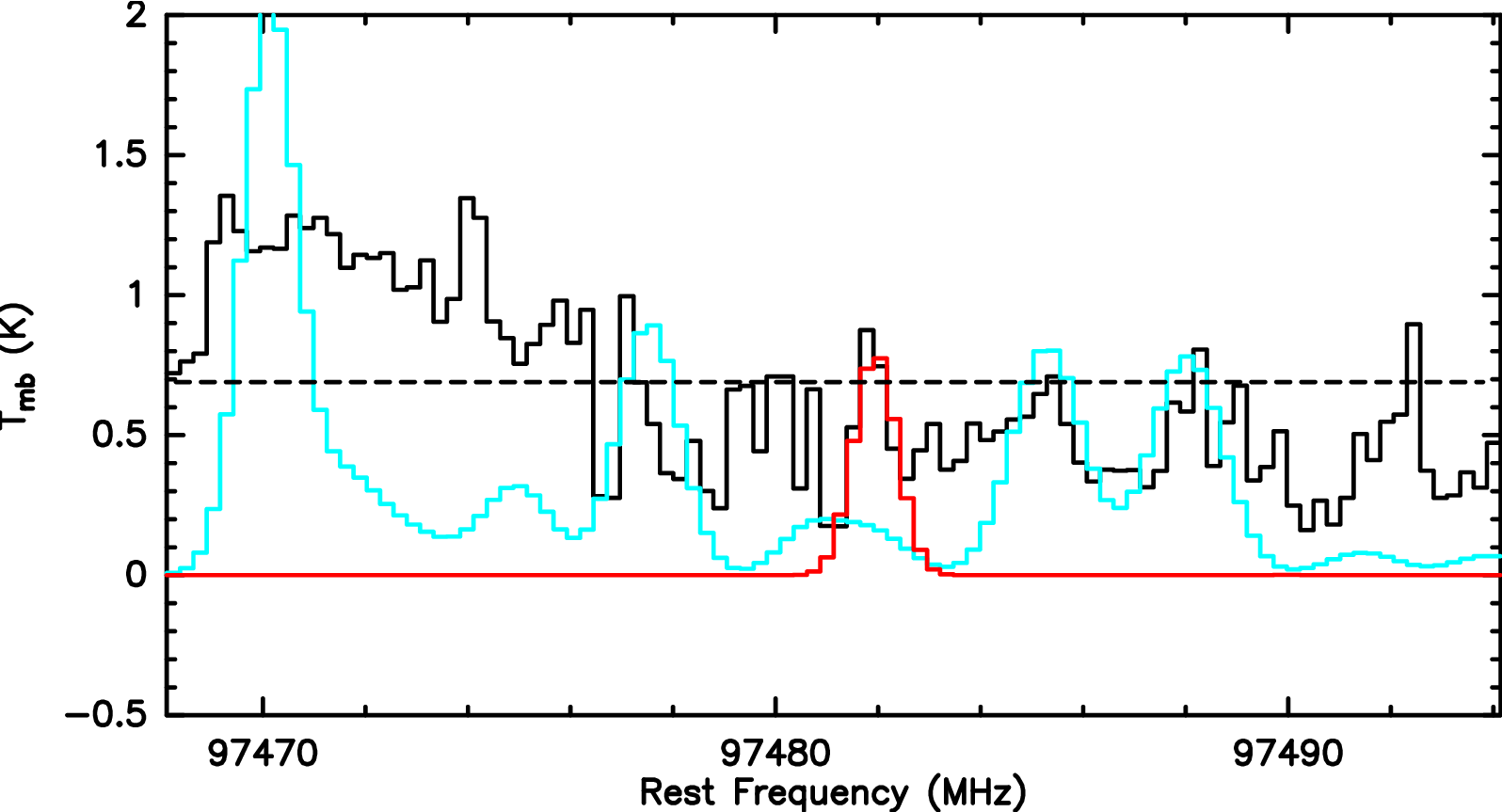}
\includegraphics[width=2.5in]{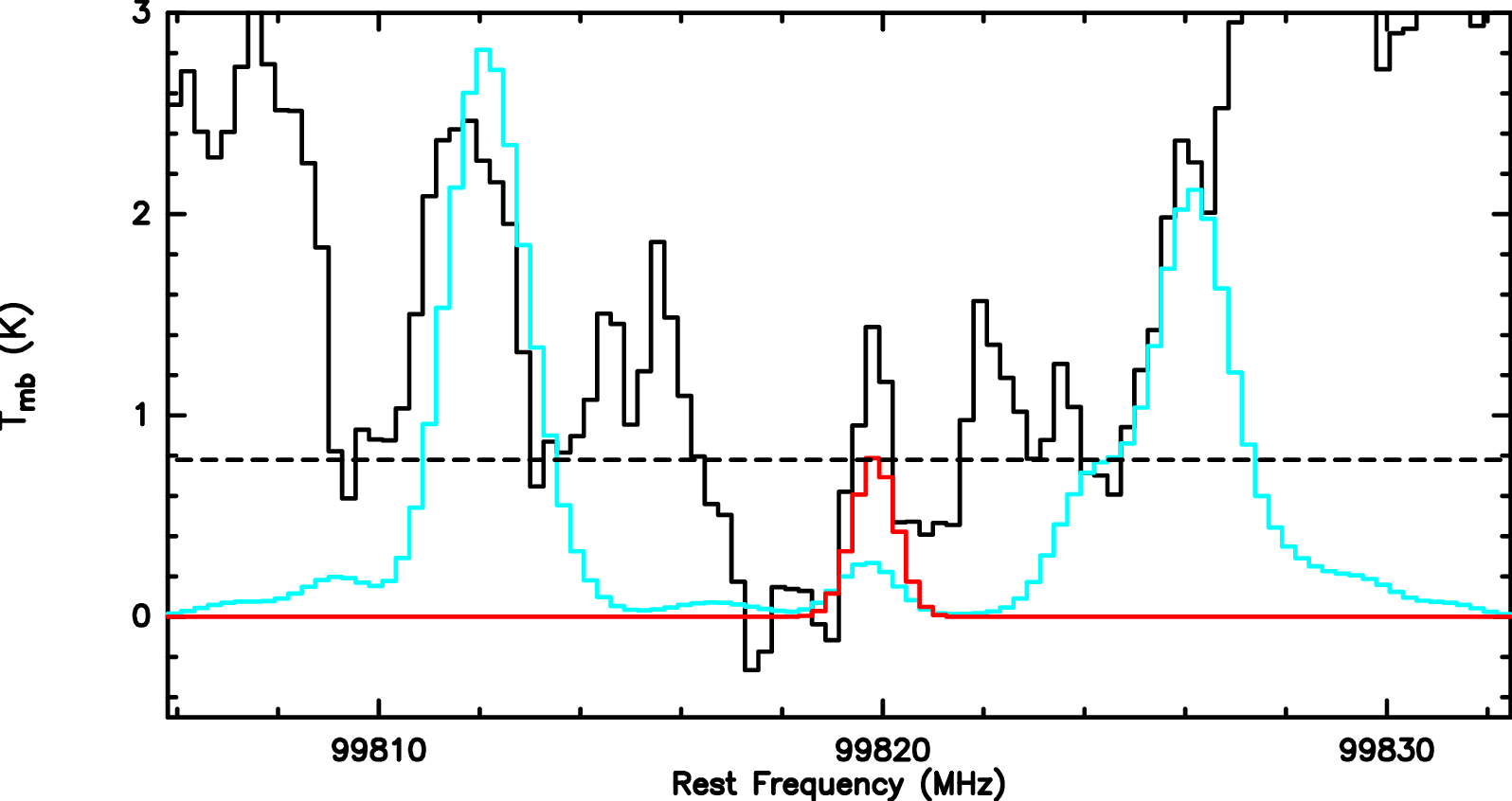}
 \includegraphics[width=2.5in]{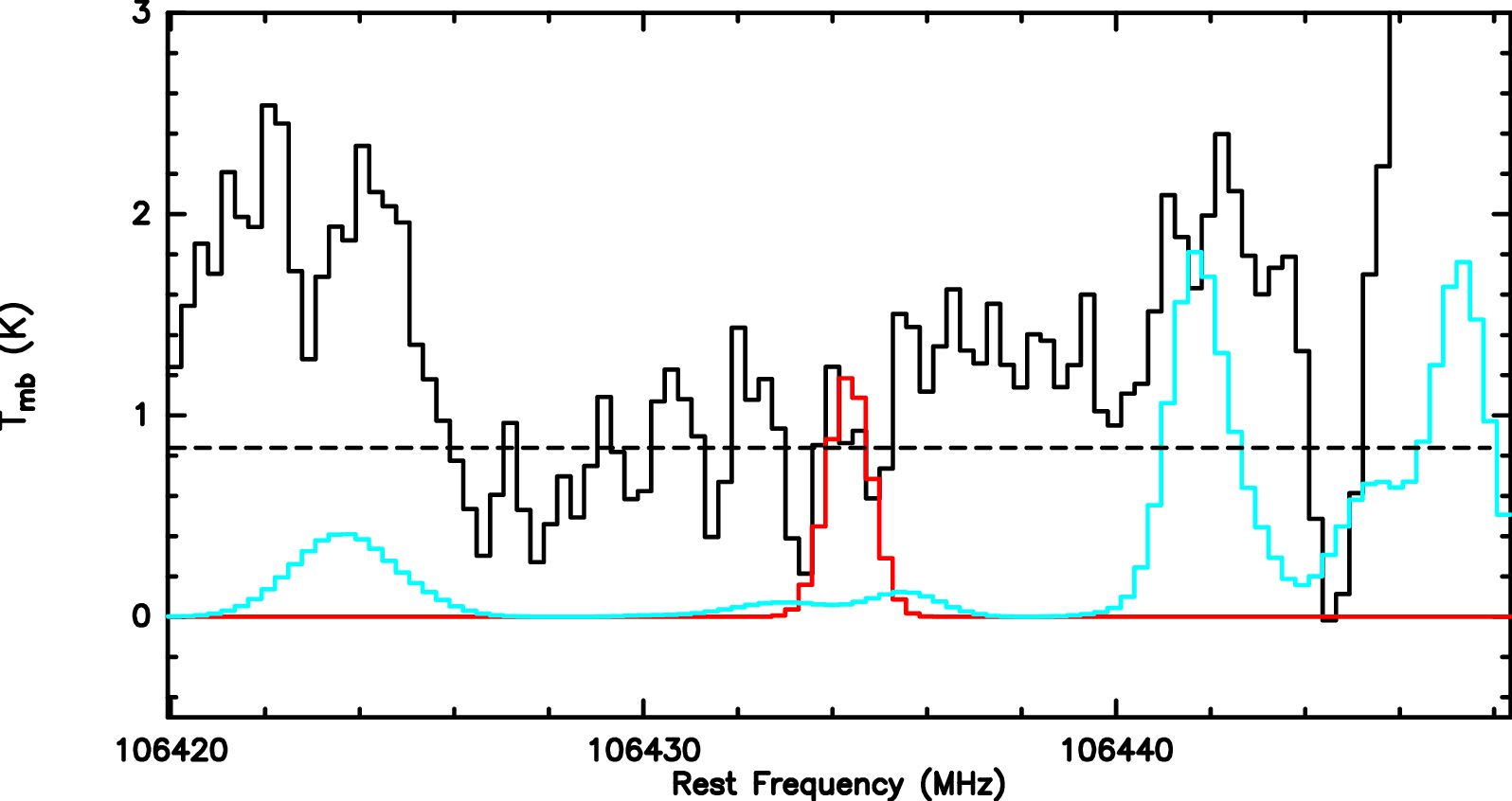}
  \includegraphics[width=2.5in]{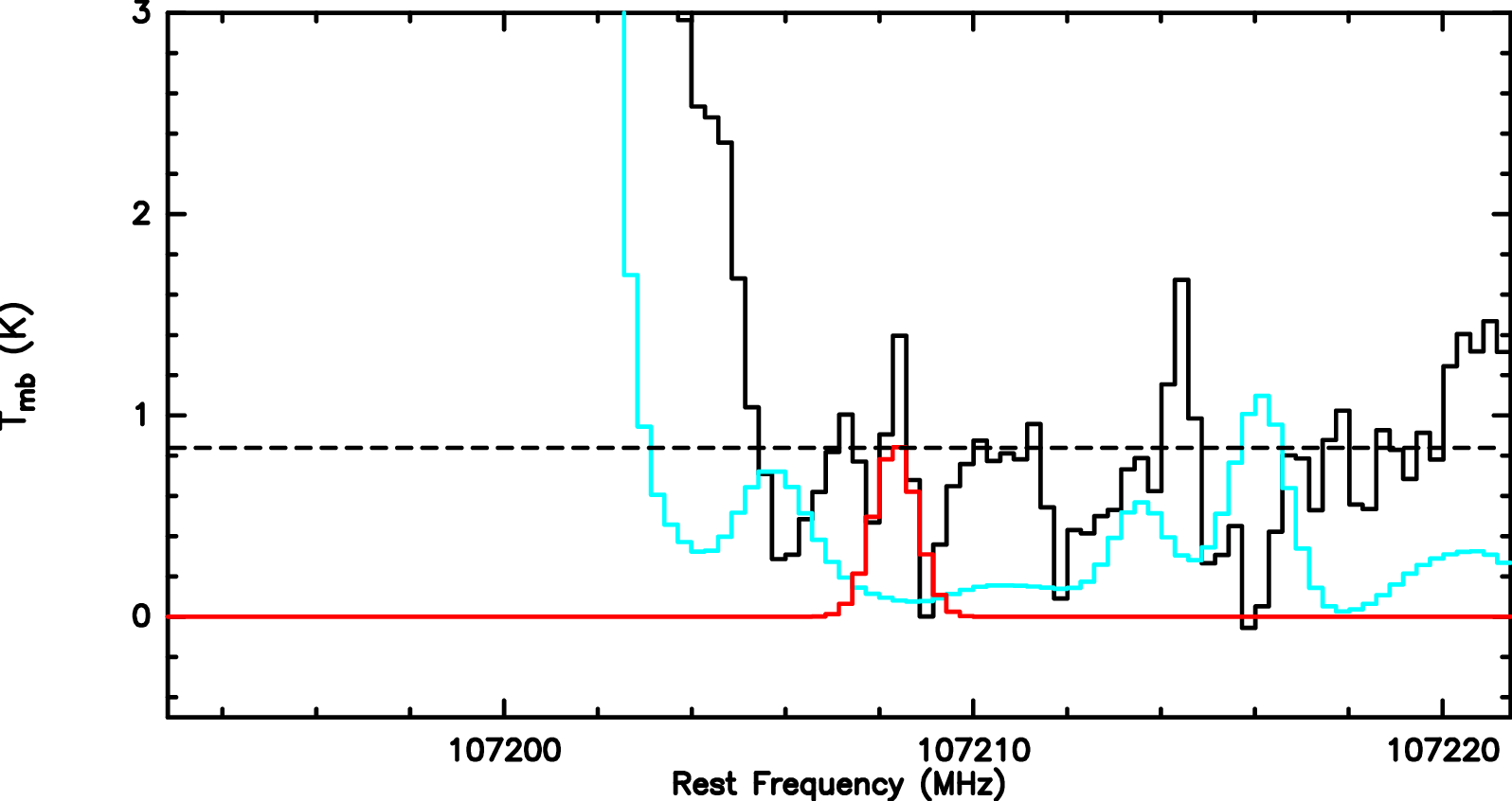}
 \includegraphics[width=2.5in]{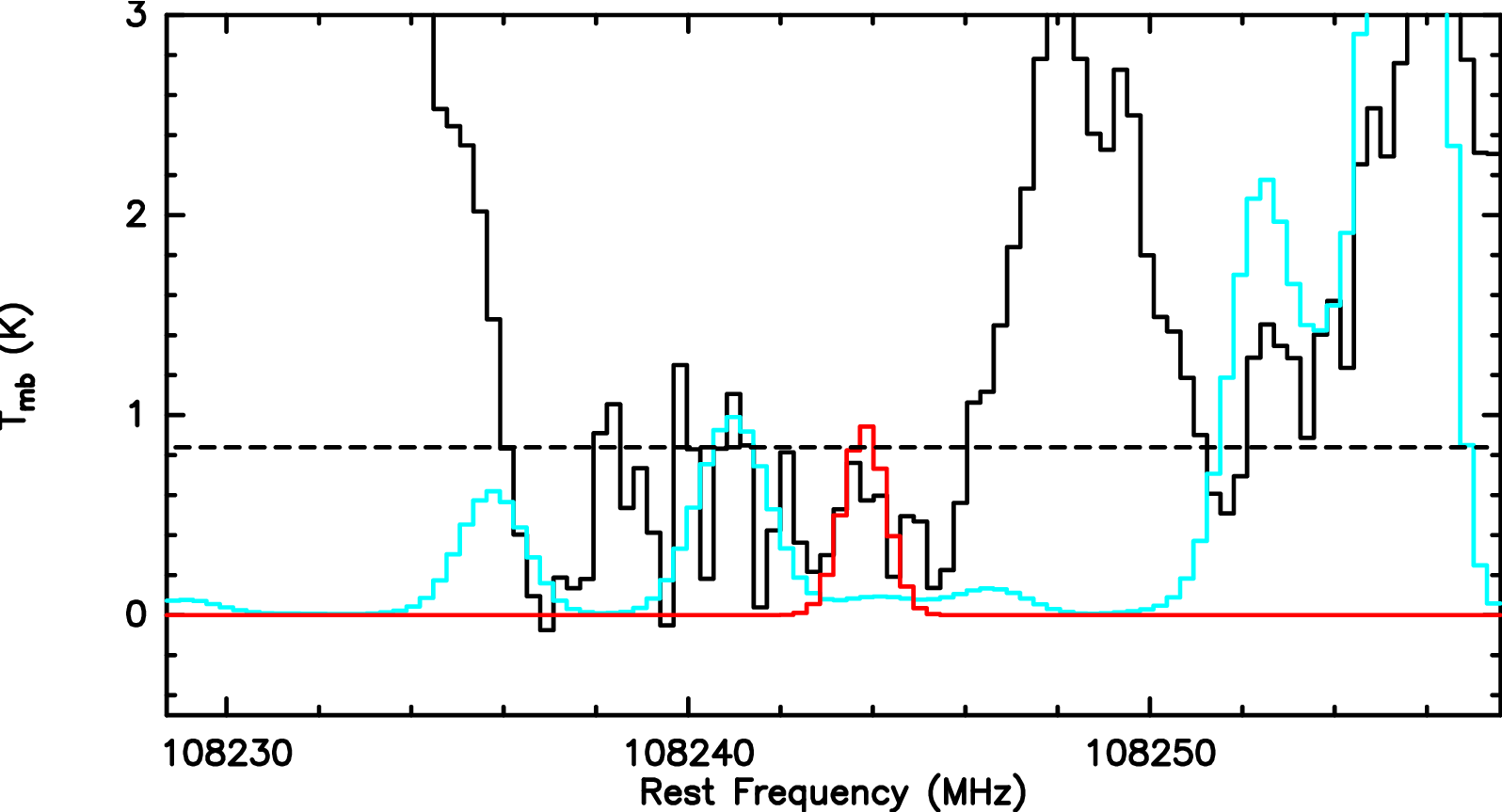}
 \includegraphics[width=2.5in]{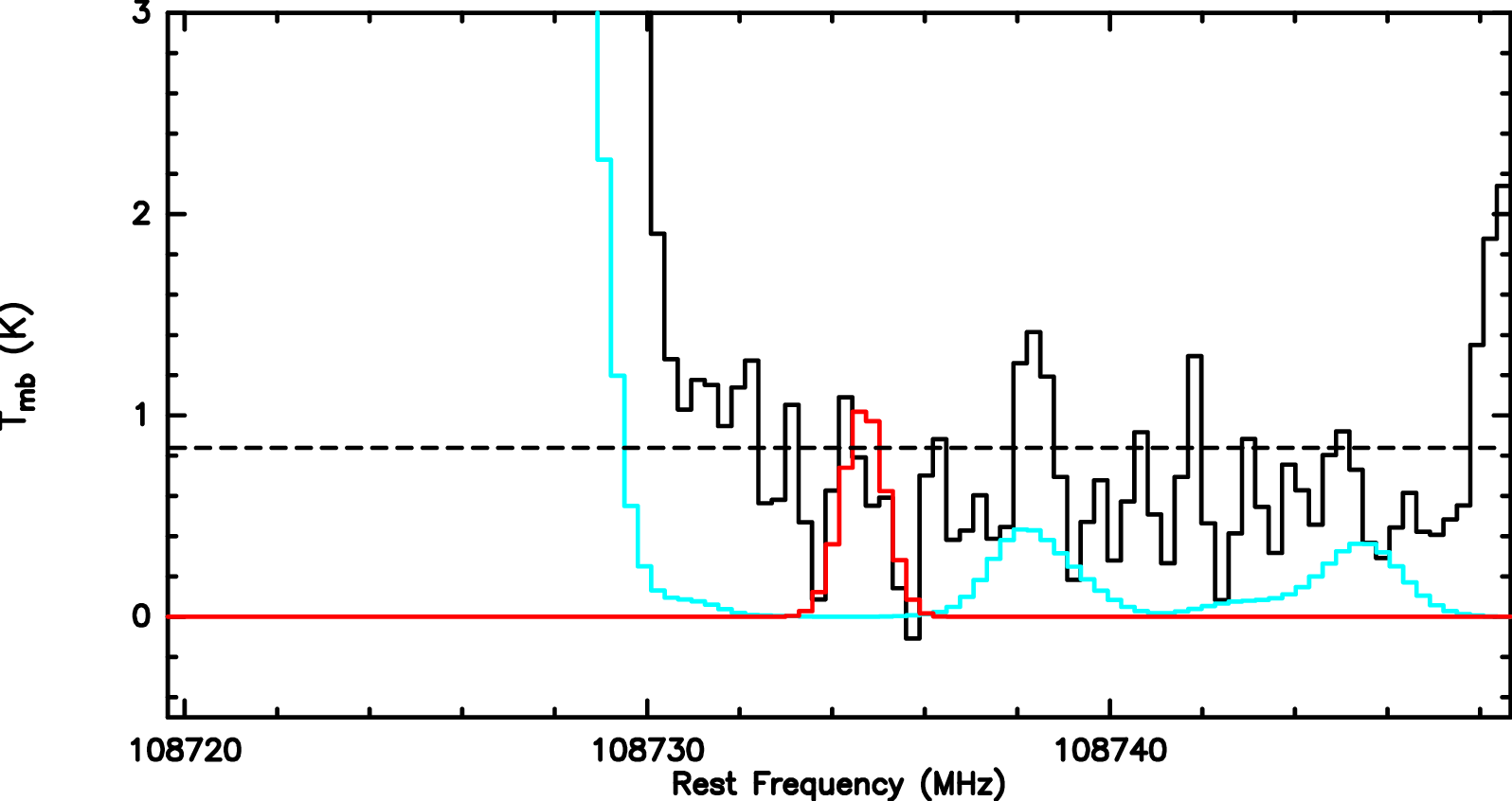}
}
\caption{Unblended transitions of NCCONH$_2$ toward Sgr~B2(N1E). Black lines show continuum-subtracted spectrum observed with the ALMA telescope. The black dashed lines show the 3$\sigma$ noise levels. The median values of the rms noise levels in the channel maps were used here. The red lines show the modeling results. The cyan lines show the modeling results of other molecules.}
\label{f:clean}
\end{figure}

 Sgr B2(N1S) is a position that is about 1\arcsec to the south of Sgr B2(N1) (see Figure~\ref{f:sgrb2}, also see Figs. 2 and 3 in \citet{Belloche19}), while Sgr B2(N1E) is about 1.5\arcsec\ to the east of Sgr B2(N1) (see Figure~\ref{f:sgrb2}, also see Figure 2 in \citet{Li21}). As Sgr B2(N1E) is on the edge of hot core, the excitation temperature and column density should be lower than those toward Sgr B2(N1S). 
 
 All the peptide-like molecules detected toward Sgr B2(N1S) were also detected toward Sgr B2(N1E). Table~\ref{tab:column} presents the column densities obtained for peptide-like molecules toward Sgr B2(N1E). We compared these results with those of Sgr B2(N1S) presented in \citet{Belloche19}. The abundances of NH$_2$CHO, CH$_3$NCO and CH$_3$NHCHO toward Sgr B2(N1E) were about one tenth of those toward Sgr B2(N1S). The abundance ratio of CH$_3$NCO to NH$_2$CHO, was 0.086 toward Sgr B2(N1E), which is in agreement with that of Sgr B2(N1S) \citep{Belloche19}. However, the abundances of CH$_3$CONH$_2$ toward Sgr B2(N1E) was only one twentieth of those toward Sgr B2(N1S). A possible explanation for the low abundance of CH$_3$CONH$_2$ and undetection of NH$_2$CONH$_2$ is that the desorption energy required by CH$_3$CONH$_2$ and NH$_2$CONH$_2$ are higher in comparison with NH$_2$CHO, CH$_3$NCO and CH$_3$NHCHO. According to laboratory studies in \citet{Ligterink18}, the desorption peak of NH$_2$CONH$_2$ is $\sim$265 K, while the desorption peak of NH$_2$CHO is $\sim$210 K, and the desorption peak of CH$_3$CONH$_2$ is $\sim$219 K. In this case, both CH$_3$CONH$_2$ and NH$_2$CONH$_2$ may not efficiently desorb into gas phase, which is consistent with results present here. It is noted that the derived rotational temperature of NH$_2$CONH$_2$ toward Sg B2(N1S) seems to be higher than other molecules (see Table 5 in \citet{Belloche19}), while the rotational temperature of CH$_3$CONH$_2$ is also higher than NH$_2$CHO, CH$_3$NCO and CH$_3$NHCHO \citep{Belloche19}. We did not search for C$_2$H$_5$CONH$_2$ and NCCONH$_2$ toward Sgr B2(N1S) because of serious line blending from other molecules.

\subsection{Possible formation mechanism for interstellar cyanoformamide}
\label{possible}


The formation of NCCONH$_2$ in the ISM is yet unclear. \citet{Winnewisser05} proposed that \ncconh\ could form from the formamide and CN:
\begin{equation}
CN + NH_2CHO \longrightarrow NCCONH_2 + H.
\end{equation}
We ran quantum chemical calculations to further study this reaction. The structures of all the species studied in this work (reactants, intermediates, transition states, and products) were first optimized in the framework of the density functional theory (DFT) employing the M06-2X \citep{zhao08} in conjunction with the 6-311+G(d,p) basis set \citep{ditchfield71}, from which rotational constants, harmonic vibrational frequencies, and zero-point energies(ZPEs) were obtained. High-performance single point energies were also calculated at the M06-2X/aug-cc-pVTZ \citep{dunning89} level using the M06-2X structures. All quantum chemical calculations were run with the Gaussian 16 \citep{frisch16}.

NH$_2$CHO molecule has several spots that could be attacked by the CN radical: the N, C, and O atoms. Therefore, besides Reaction (1), the following reactions might be also possible. Thus, we ran quantum mechanical calculations to evaluate these two reactions.
\begin{equation}
CN + NH_2CHO \longrightarrow NH_2CN + HCO,  
\end{equation}
\begin{equation}
CN + NH_2CHO \longrightarrow trans$-$/cis$-$NCOCHNH + H. 
\end{equation}
A schematic of the full PES for the reaction between CN and NH$_2$CHO can be seen in Figure \ref{f:energy}. Reaction (1) may initially proceed via the formation of intermediate I1 (NH$_2$C(O)HCN) via transition state TS1. Then the intermediate II is linked to the formation of the products P1 (NH$_2$COCN + H) via transition state TS2, as this involves the elimination of an H-atom. The reaction is exothermic. The barrier energy of this reaction is equal to the energy of transition state minus the energy of the two reactants, which is 9.7 kJ/mol (=1,167 K). We have calculated the rate coefficients of Reaction (1) by the modified Arrhenius equation. The rate coefficients are $ 2.4 \times 10^{-13}$ and $1.94 \times 10^{-12}$ cm$^3$ s$^{-1}$ at T = 150 and 200 K, respectively. This means that the exothermic formation reaction (1) is slow in the gas-phase. Reaction (2) is via transition state TS3 to form products P2 (NH$_2$CN + HCO), and Reaction (3) firstly form the intermediate I2 (NH$_2$C(H)OCN), then via transition state TS4 and TS5 to form two products P3 (cis-NCOCHNH + H) and P4 (trans-NCOCHNH + H), respectively. As Figure  \ref{f:energy} shows,  both reactions (2) and (3) are endothermic with high barriers ($>$55 kJ/mol). Their rates would be very low under the ISM conditions.

Besides, we propose another possible reaction route of NCCONH$_2$, where it might be formed upon recombination of two free radicals on the grain surface at low temperature as the following:
\begin{equation}
CN + NH_2CO \longrightarrow NCCONH_2. 
\end{equation}
The radical NH$_2$CO can be formed by CN with water molecules of the ice mantle as reported by \citet{Rimola18}.

\section{CONCLUSIONS}\label{sec:conclusions}

In this paper, we present the tentative detection of \ncconh\ in ISM for the first time. The main results are summarized as follows:

1. Ten unblended transitions, and twenty one partially-blended transition of \ncconh\ were detected around 3$\sigma$ noise levels toward Sgr B2(N1E) with the ALMA archive data at 3-mm wavelength. 

2. The column density of \ncconh\ toward Sgr B2(N1E) was estimated to be $2.4\times10^{15}$ cm$^{-2}$, while the abundance relative to H$_2$ was estimated to be $6.9\times10^{-10}$. The abundance ratio of \ncconh\ to formamide is $\sim$ 0.01. 

3. Some related molecules were also searched toward Sgr B2(N1E). Both CH$_3$NCO and CH$_3$NHCHO were detected. The abundance ratio between NH$_2$CHO, CH$_3$NCO and CH$_3$NHCHO toward Sgr B2(N1E) is same to that toward Sgr B2(N1S), while the abundance of CH$_3$CONH$_2$ toward Sgr B2(N1E) was much lower than those toward Sgr B2(N1S). 

It is highly desirable to conduct confirmation observations on the existence of \ncconh\ in space.

\section*{ACKNOWLEDGEMENTS}

We thank the referee, for his/her critical reading of the paper, and very constructive suggestions and comments of our work.  JL would like to thank  Dr. Laura Colzi for suggestions about this work. JL thank Prof. Vadim Illyushin, Roman Motiyenko, Laurent Margules and Yong Zhang for helpful discussions. This work made use of the High Performance Computing Resource in the Core Facility for Advanced Research Computing at Shanghai Astronomical Observatory. This work has been supported by the National Key R\&D Program of China (No. 2022YFA1603101), the National Natural Science Foundation of China (Grant Nos. 11973075, U1931104), Chongqing Talents: Exceptional Young Talents Project (Grant No. cstc2021ycjh-bgzxm0027), and Fundamental Research Funds for the Central Universities (Grant No. 2020CDJXZ002). This paper makes use of the following ALMA data: ADS/JAO.ALMA\#2016.1.00243.S. ALMA is a partnership of ESO (representing its member states), NSF (USA) and NINS (Japan), together with NRC (Canada), MOST and ASIAA (Taiwan), and KASI (Republic of Korea), in cooperation with the Republic of Chile. The Joint ALMA Observatory is operated by ESO, AUI/NRAO and NAOJ. Data analysis was in part carried out on the open-use data analysis computer system at the Astronomy Data Center (ADC) of NAOJ. This research has made use of NASA's Astrophysics Data System.

\onecolumn

\clearpage

\clearpage

\begin{figure}
\centering
{\includegraphics[width=2.5in]{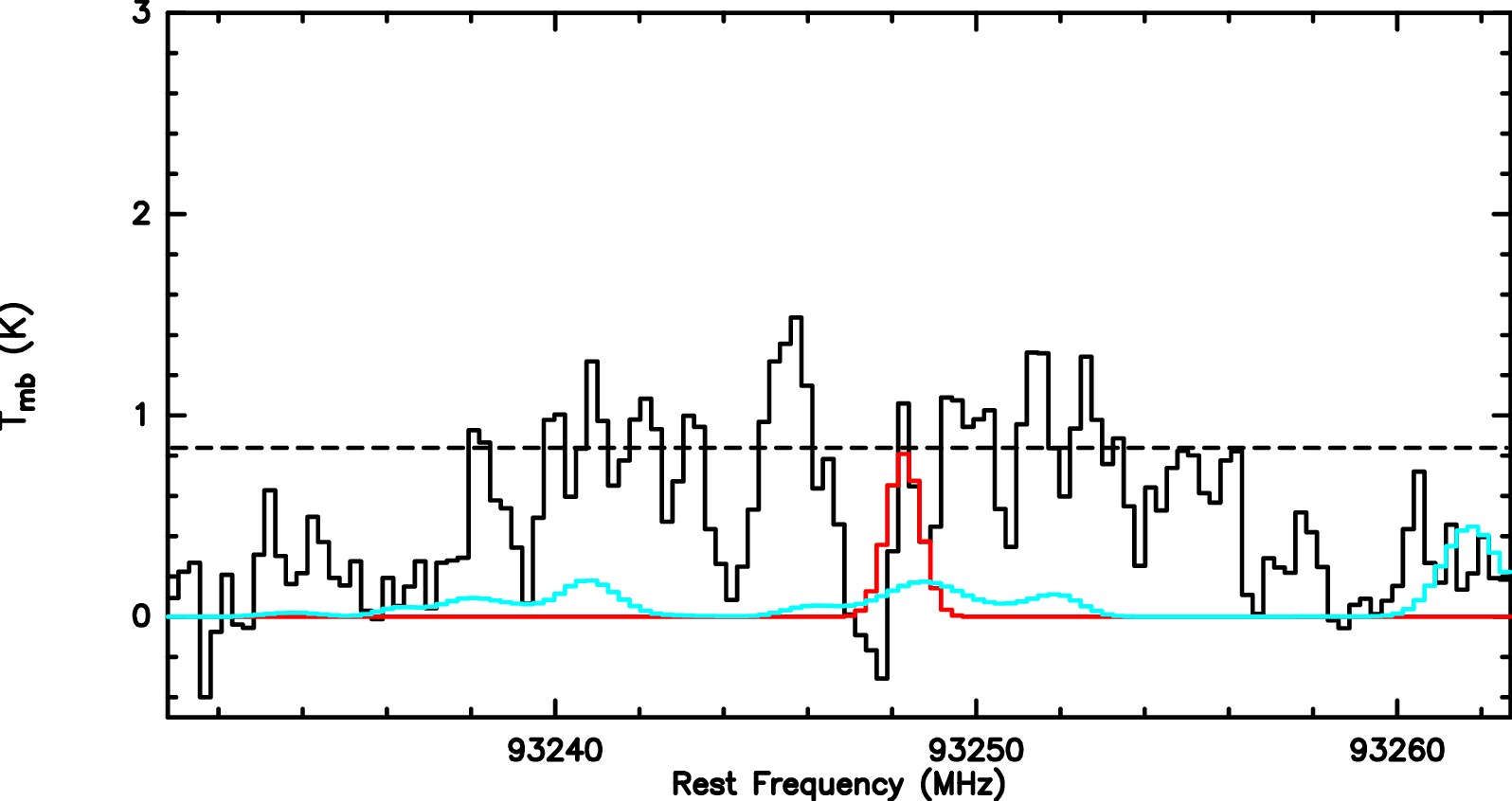}
 \includegraphics[width=2.5in]{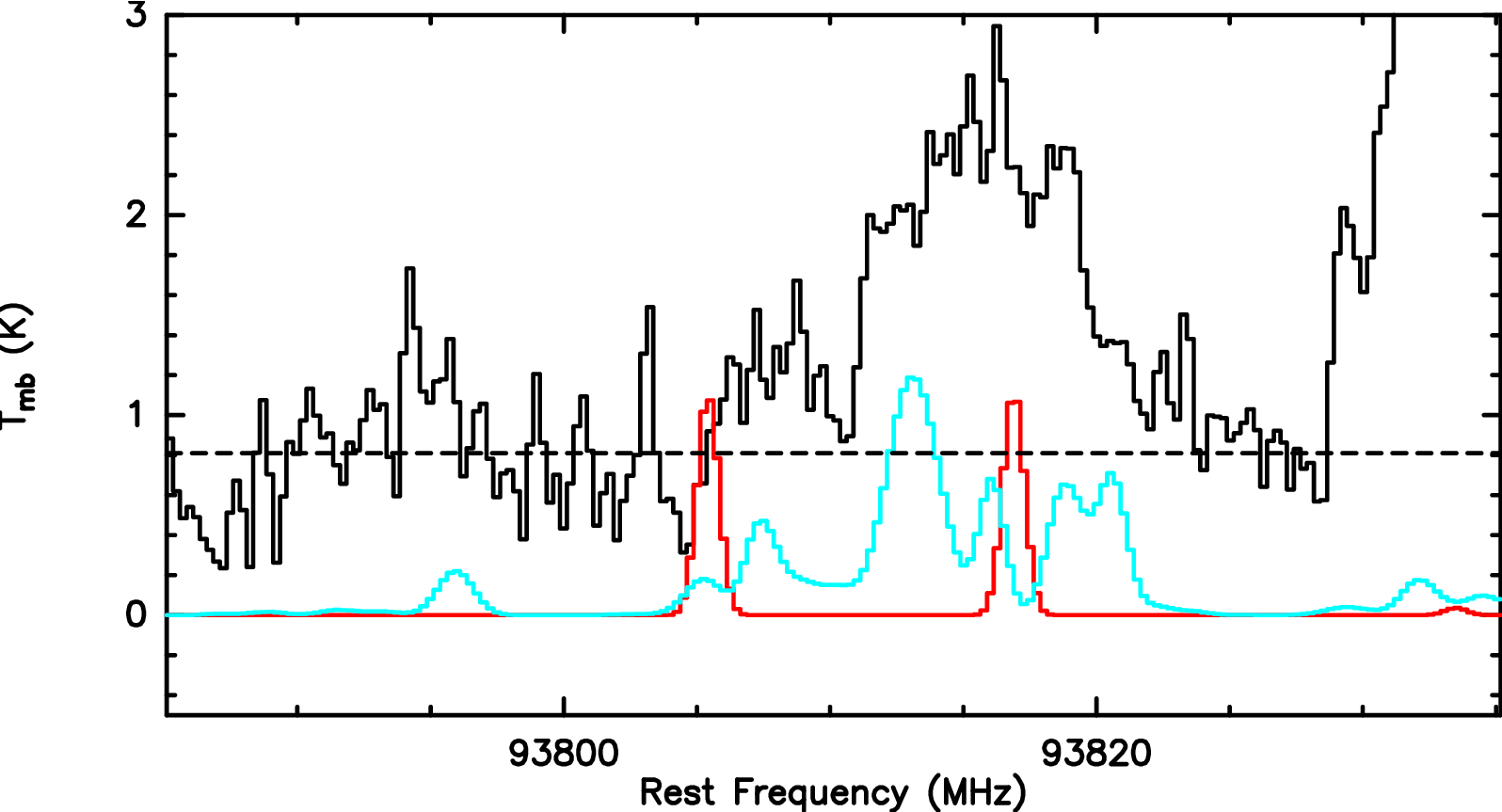}
\includegraphics[width=2.5in]{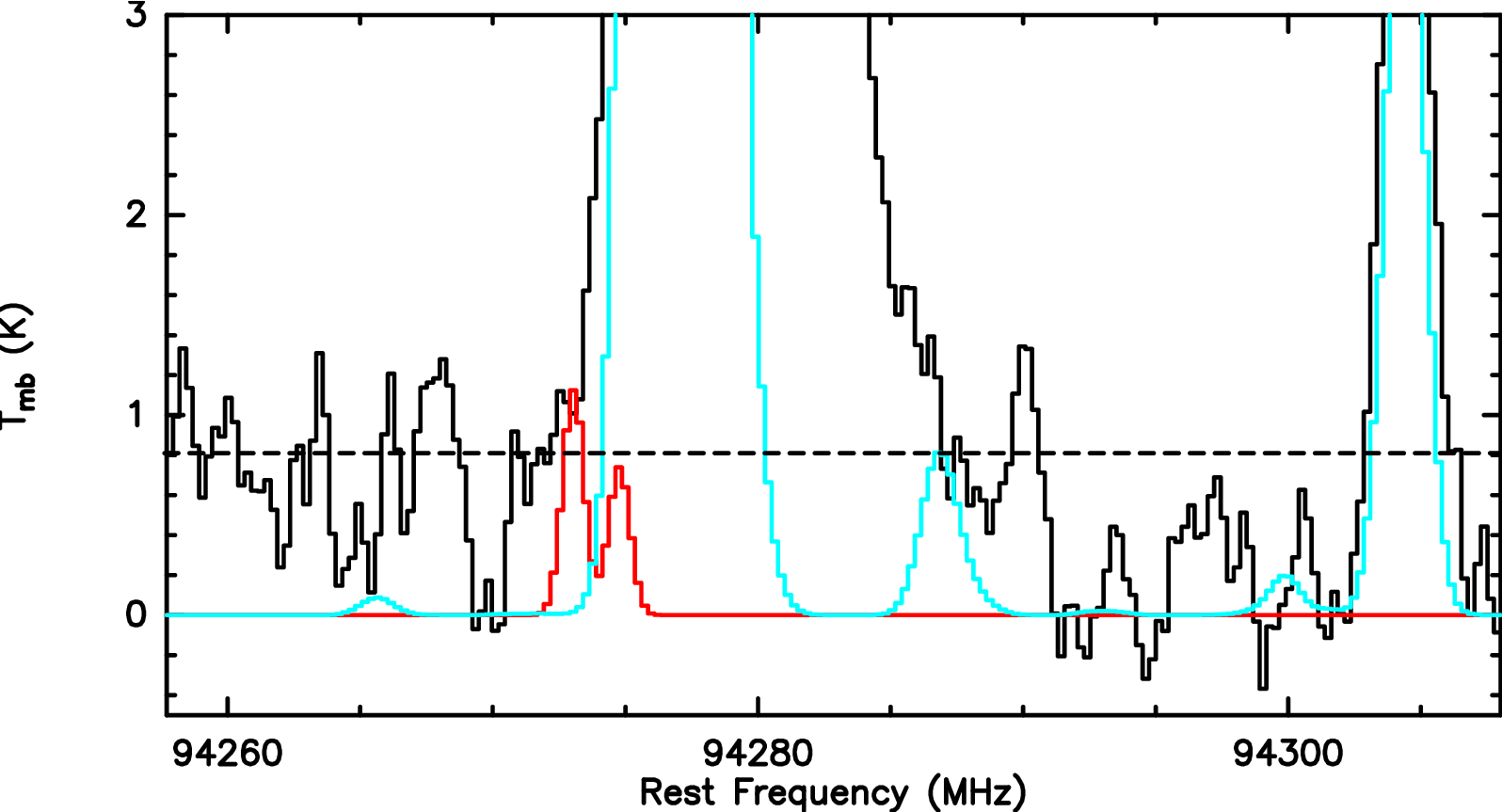}
 \includegraphics[width=2.5in]{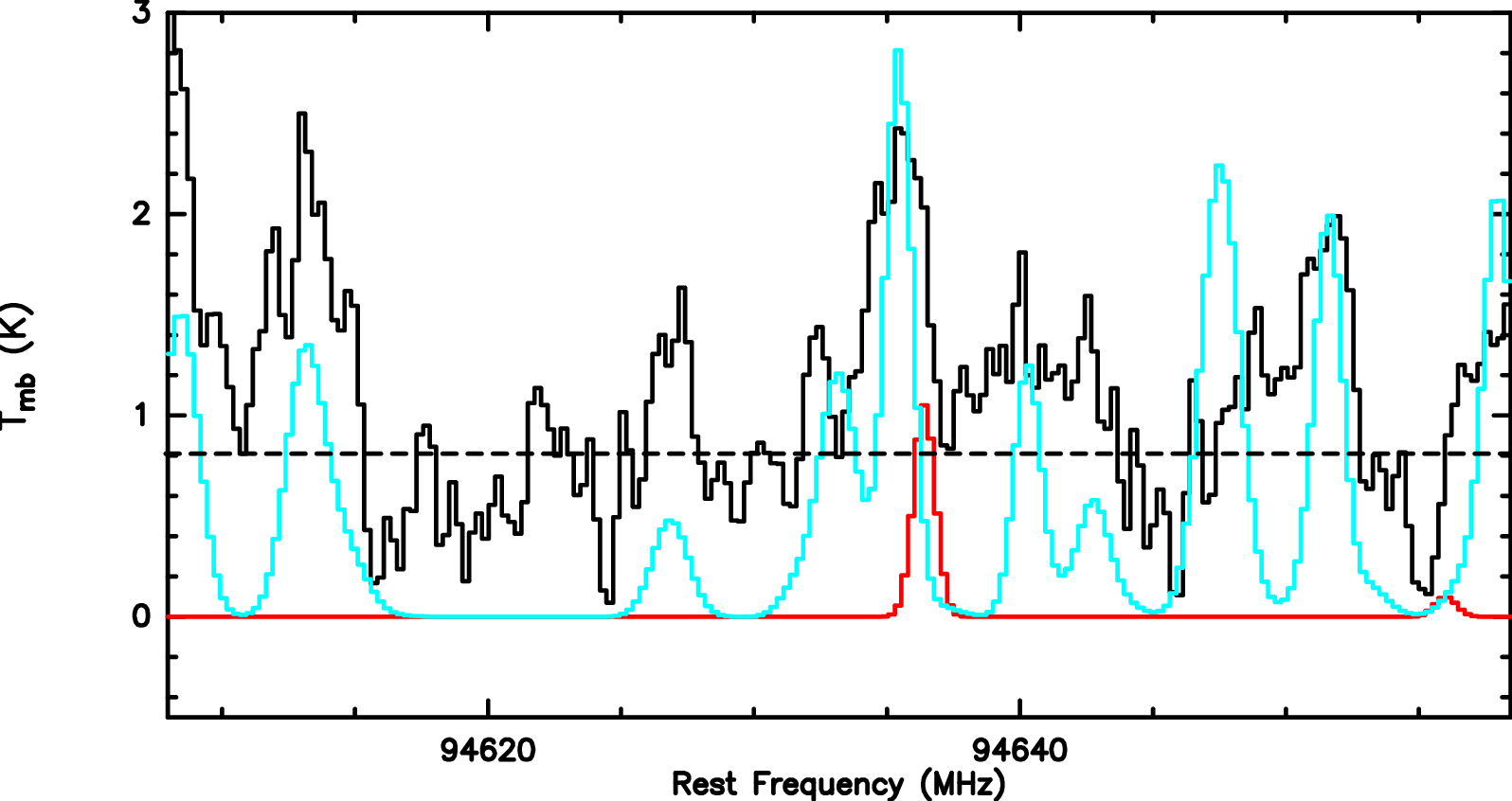} 
 \includegraphics[width=2.5in]{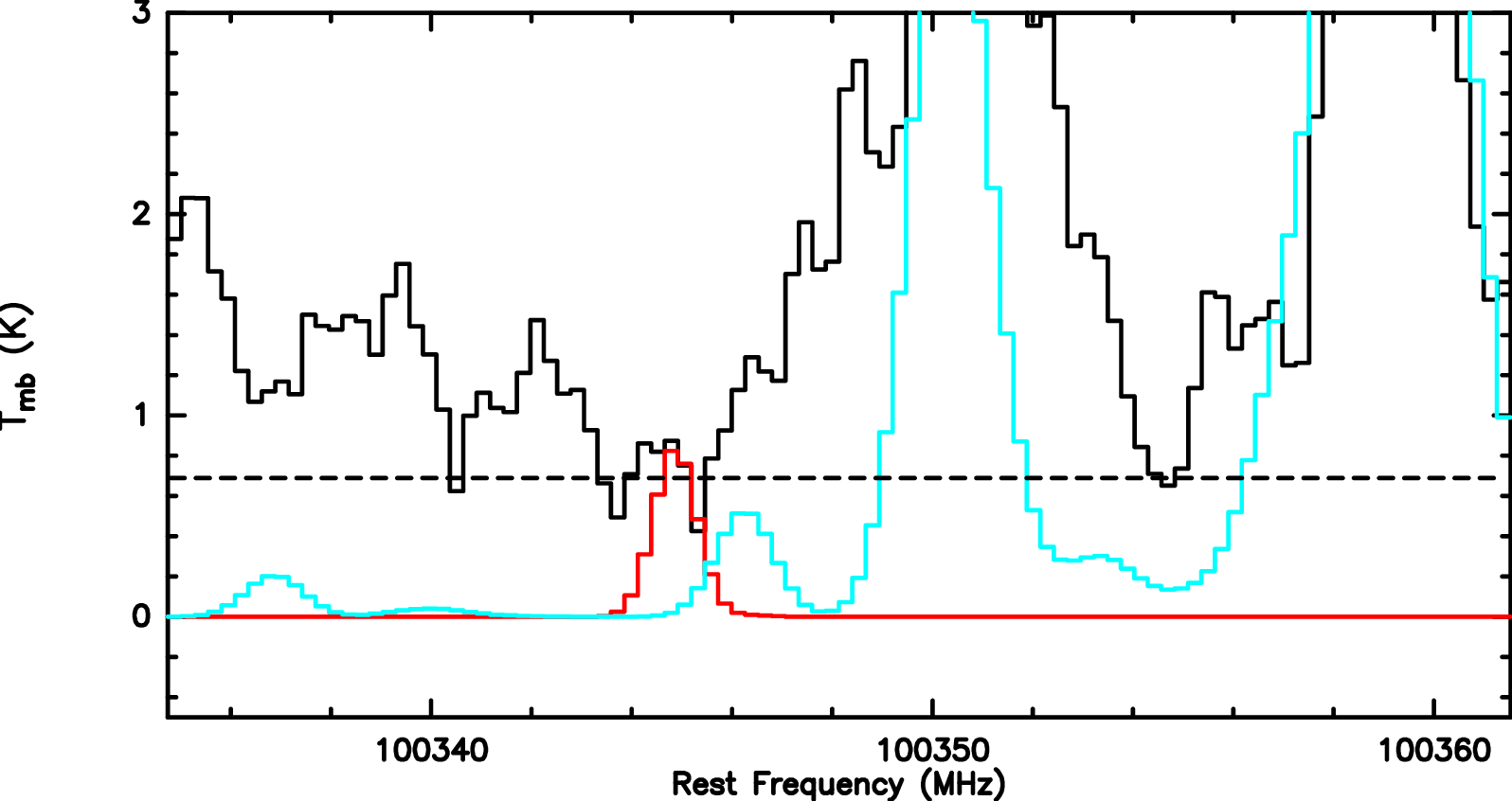}
 \includegraphics[width=2.5in]{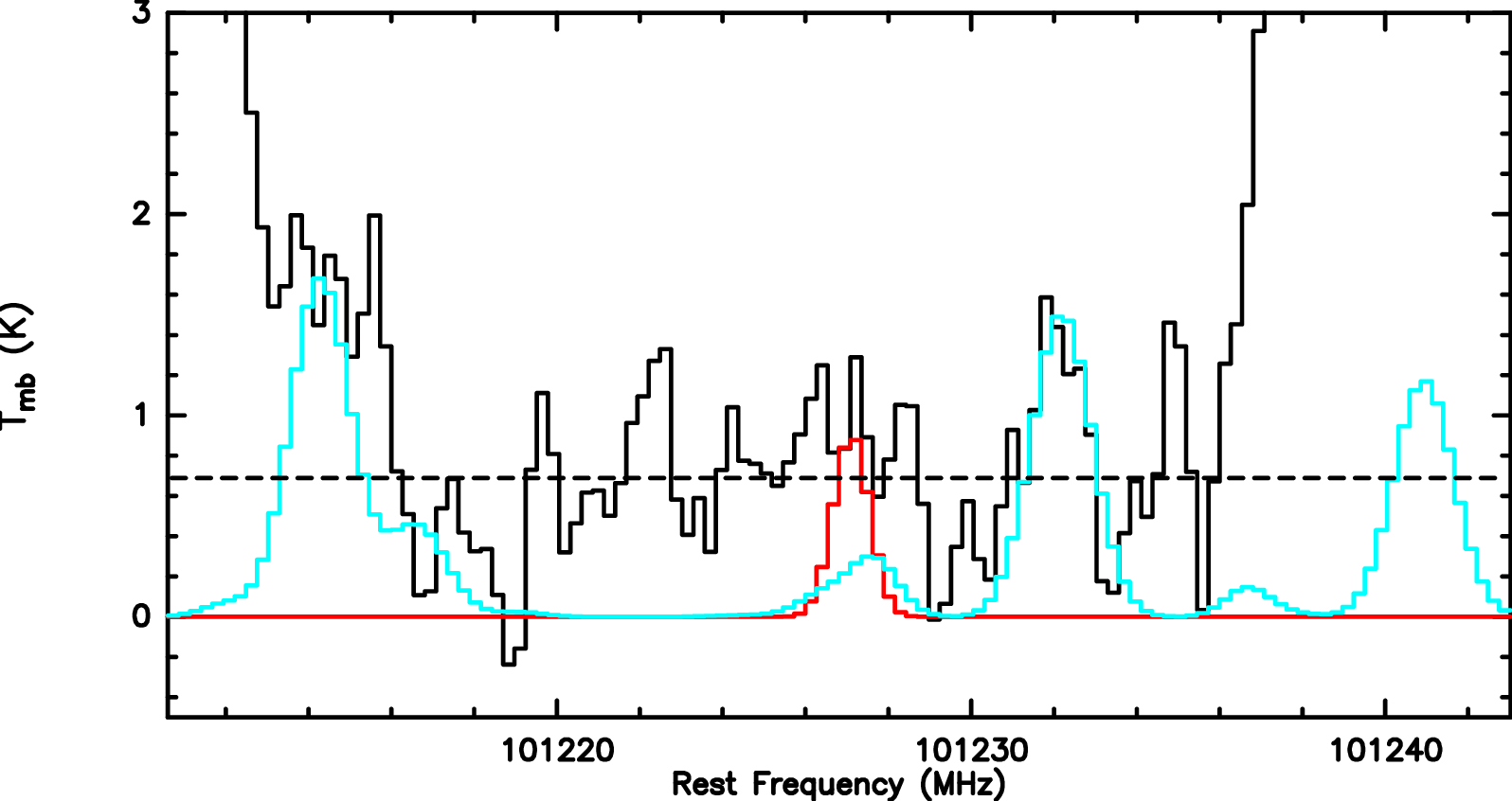}
  \includegraphics[width=2.5in]{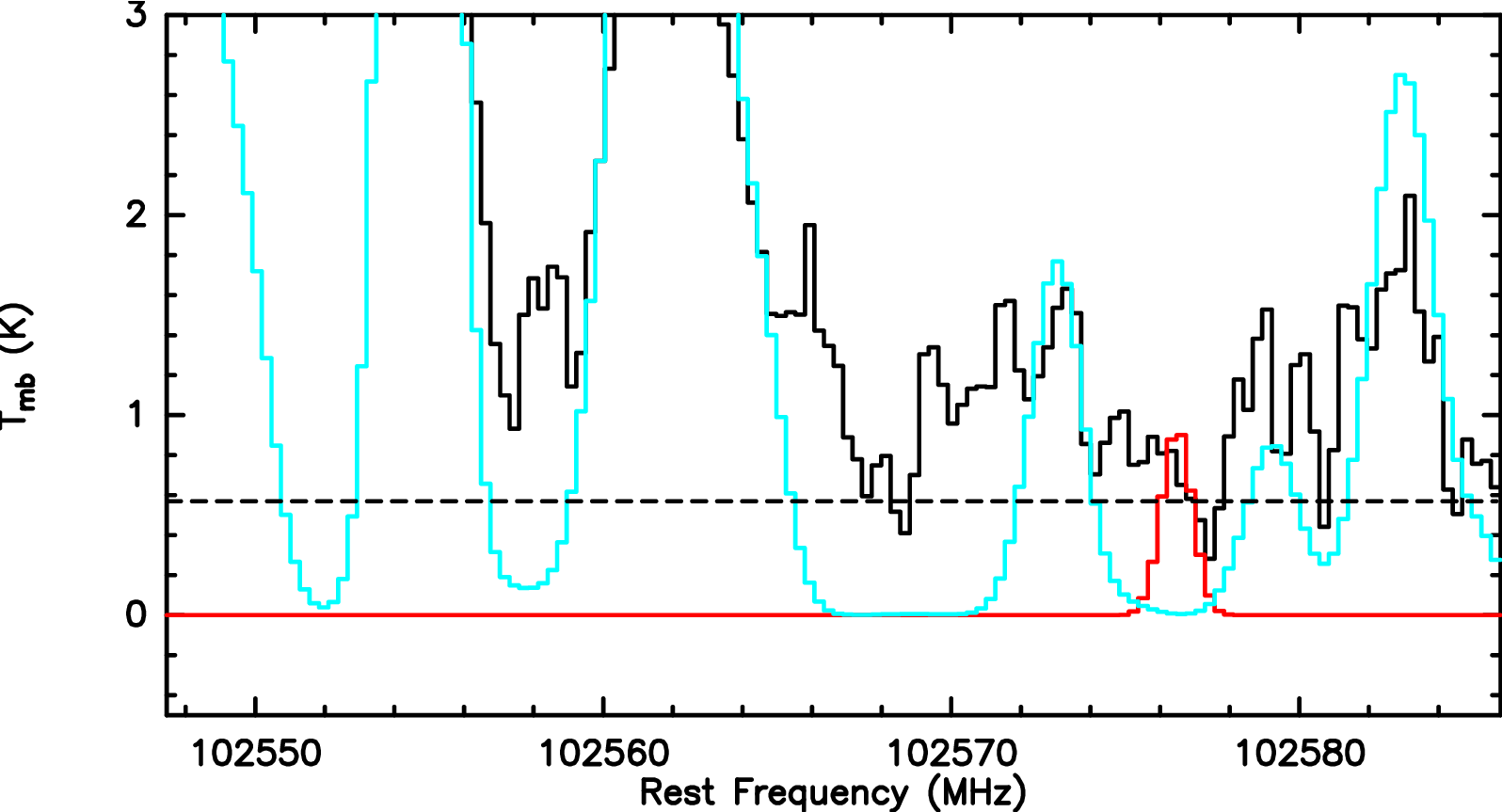}
 \includegraphics[width=2.5in]{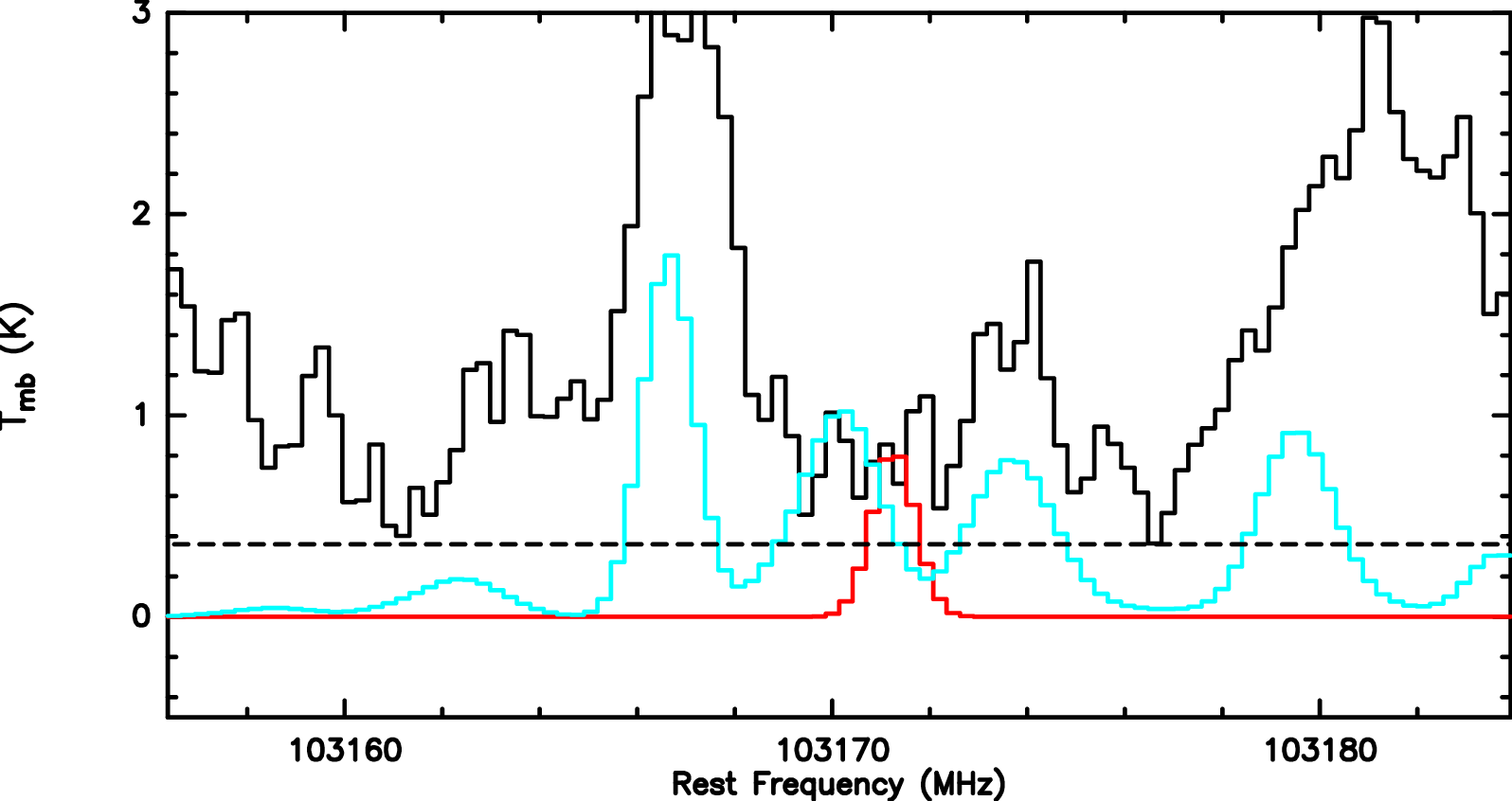}
  \includegraphics[width=2.5in]{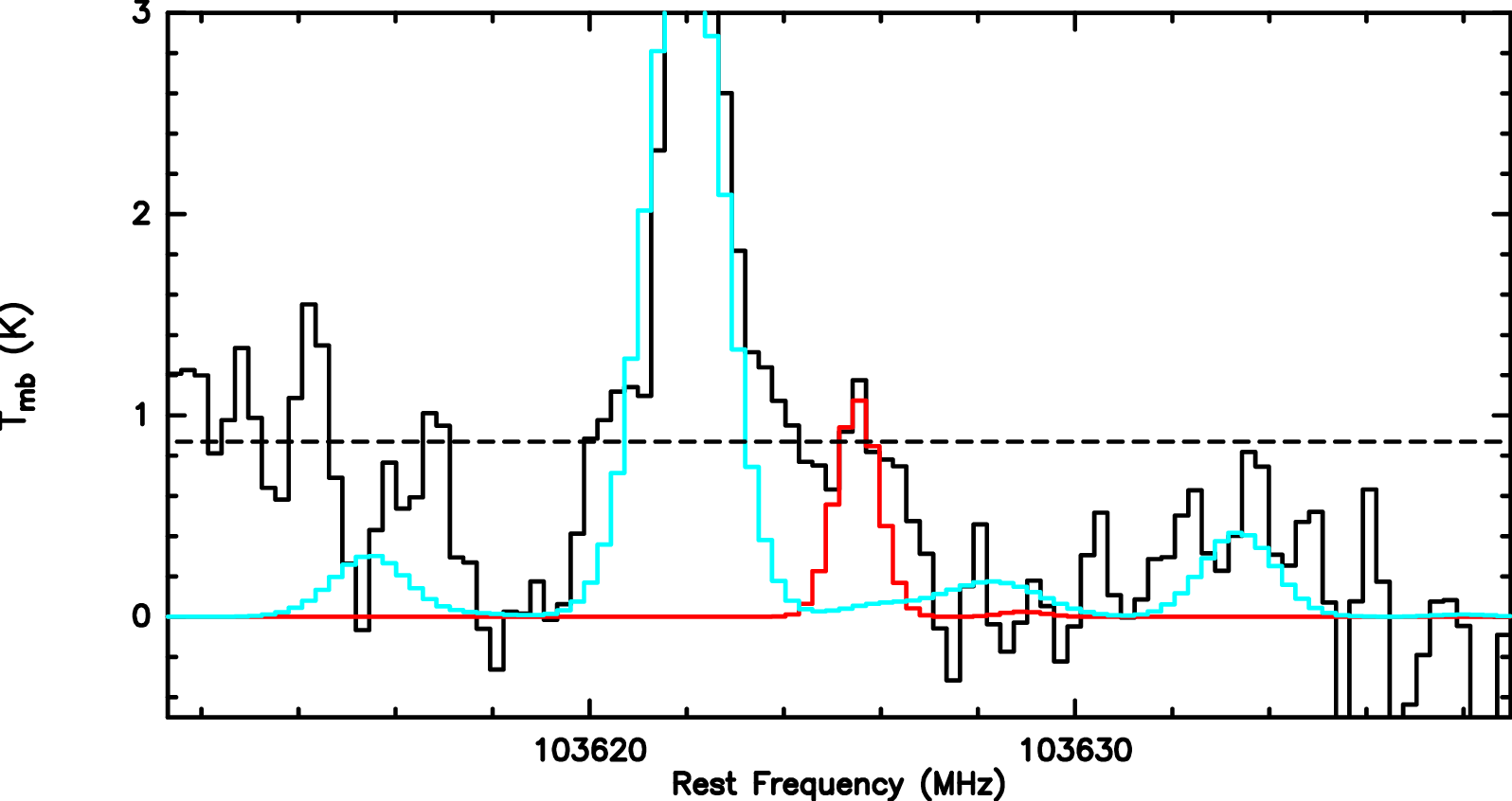}
 \includegraphics[width=2.5in]{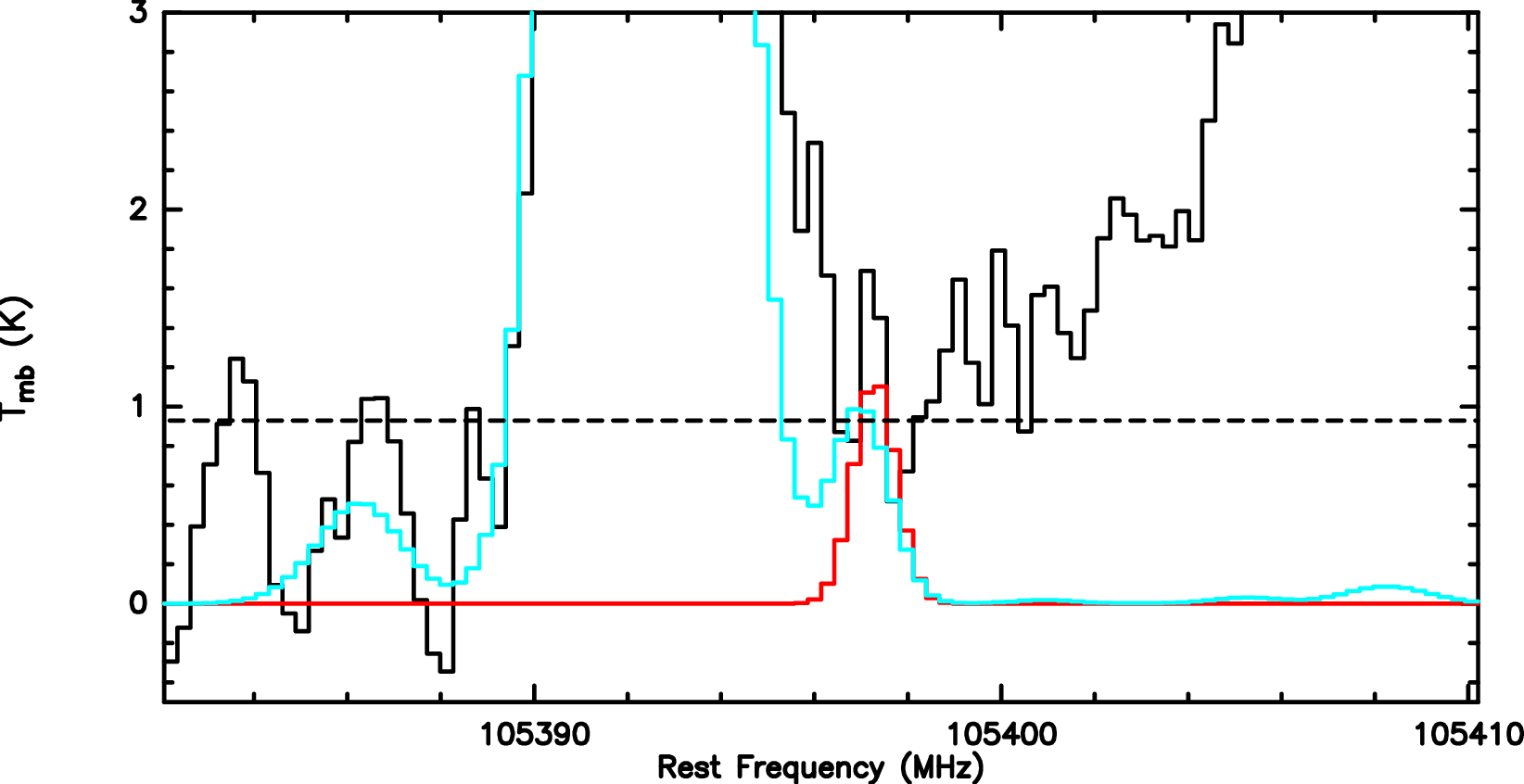}
}
\caption{Partially blended transitions of NCCONH$_2$ toward Sgr~B2(N1E). Black lines show continuum-subtracted spectrum observed with the ALMA telescope. The black dashed lines show the 3$\sigma$ noise levels derived by \citet{Belloche19}. The red lines show the modeling results. The cyan lines show the modeling results of other molecules.}
\label{f:partial}
\end{figure}

\clearpage

\begin{figure}
\addtocounter{figure}{-1}
\centering
{\includegraphics[width=2.5in]{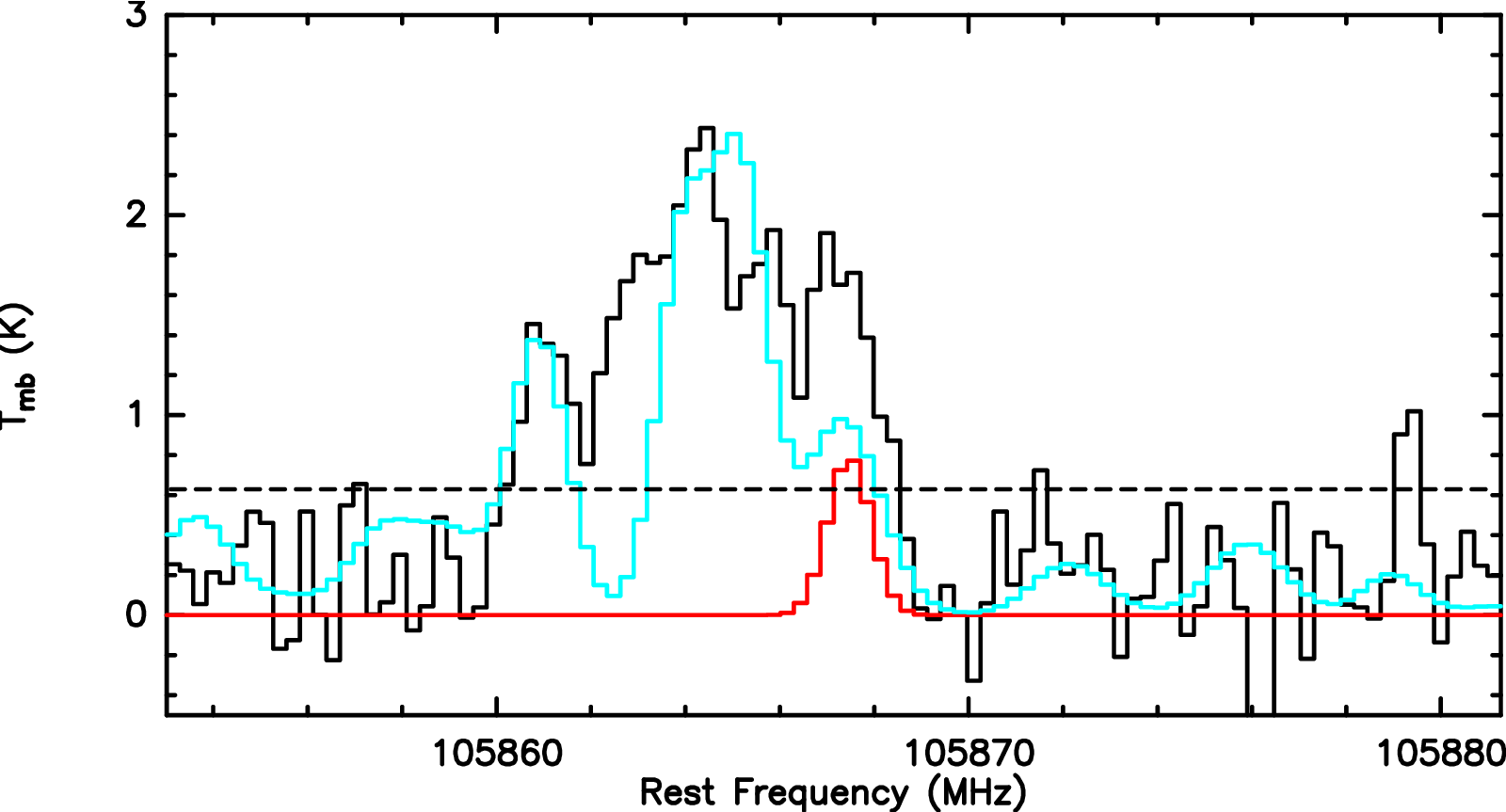}
 \includegraphics[width=2.5in]{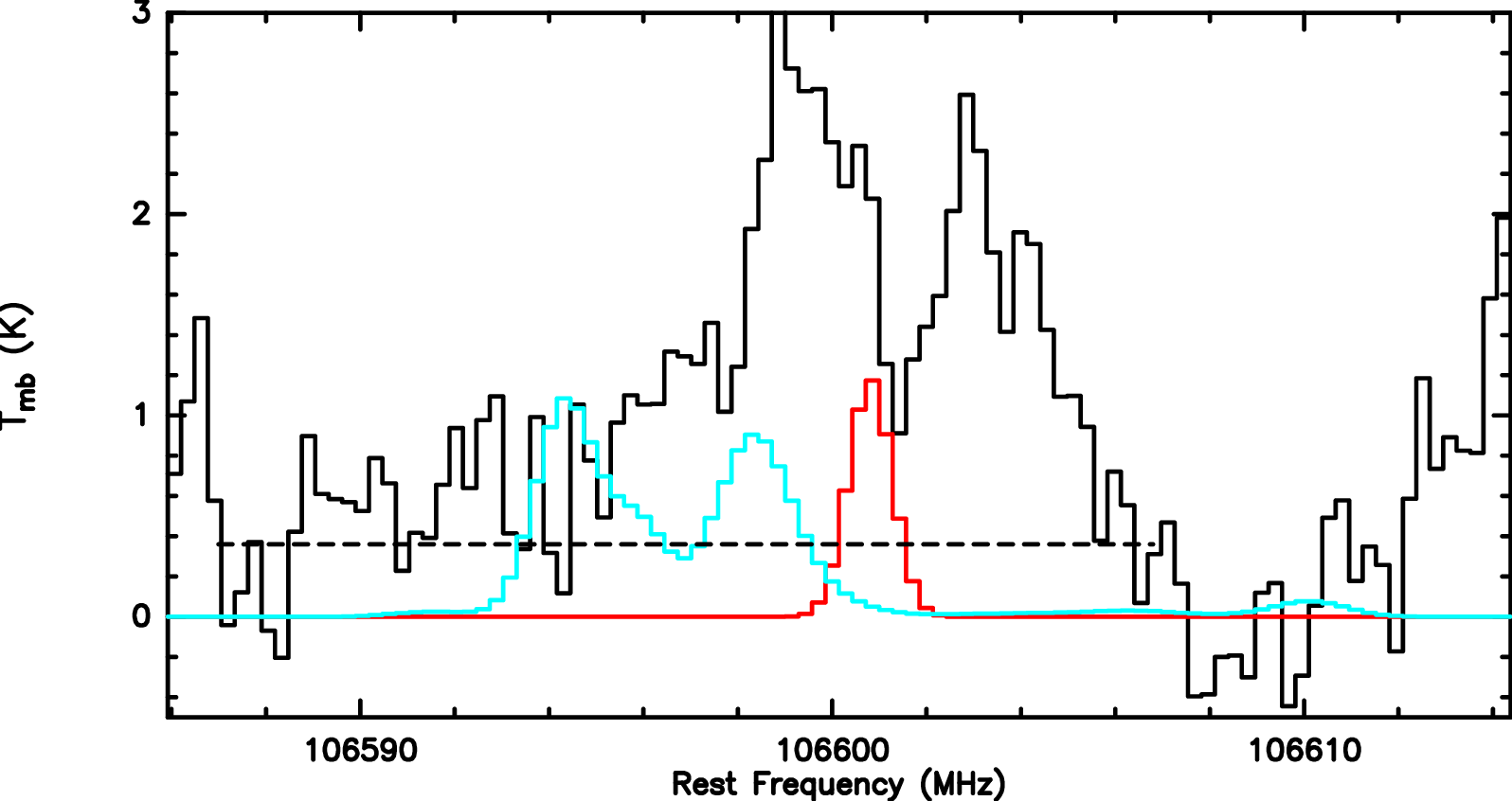}
\includegraphics[width=2.5in]{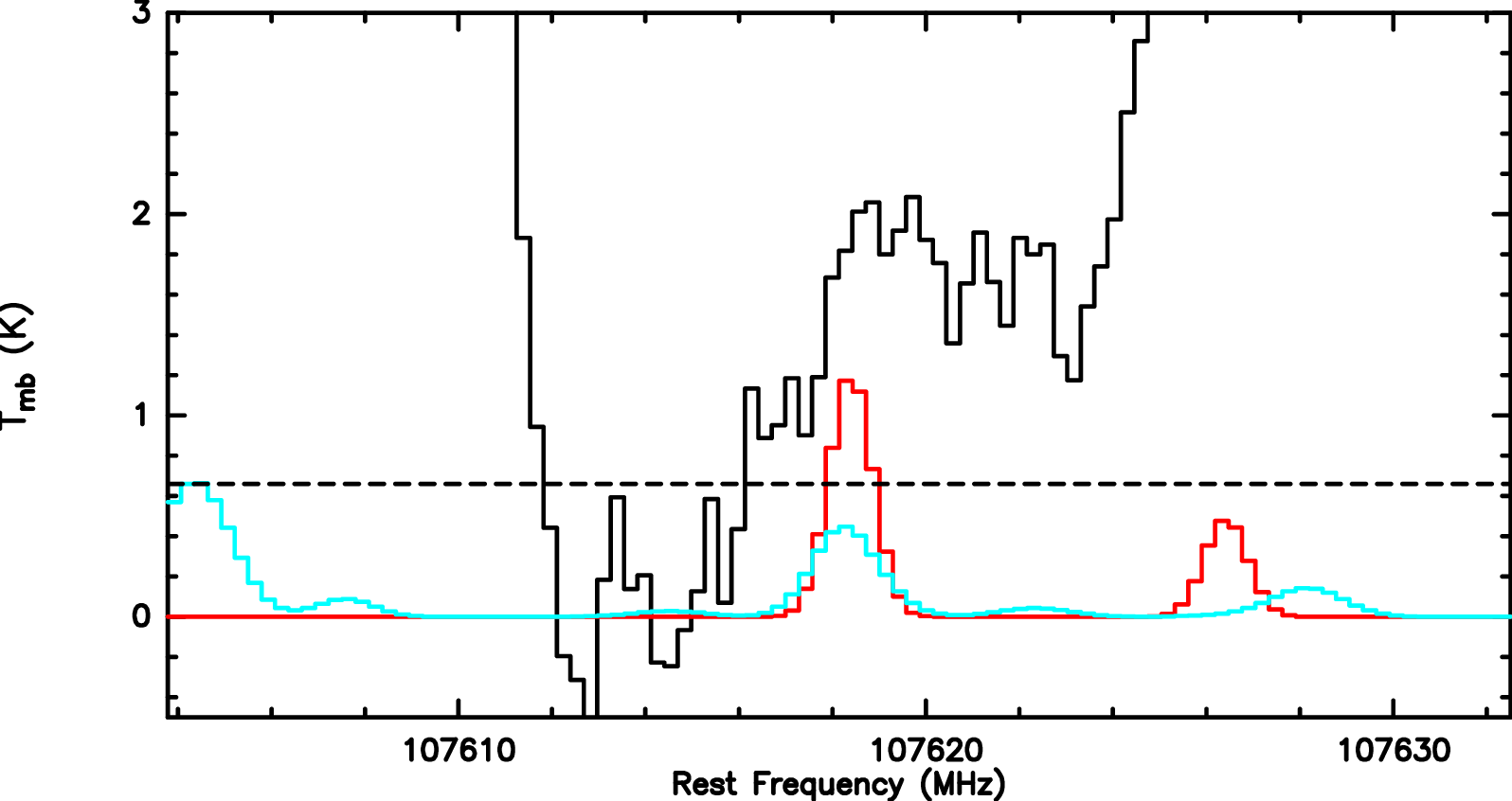}
 \includegraphics[width=2.5in]{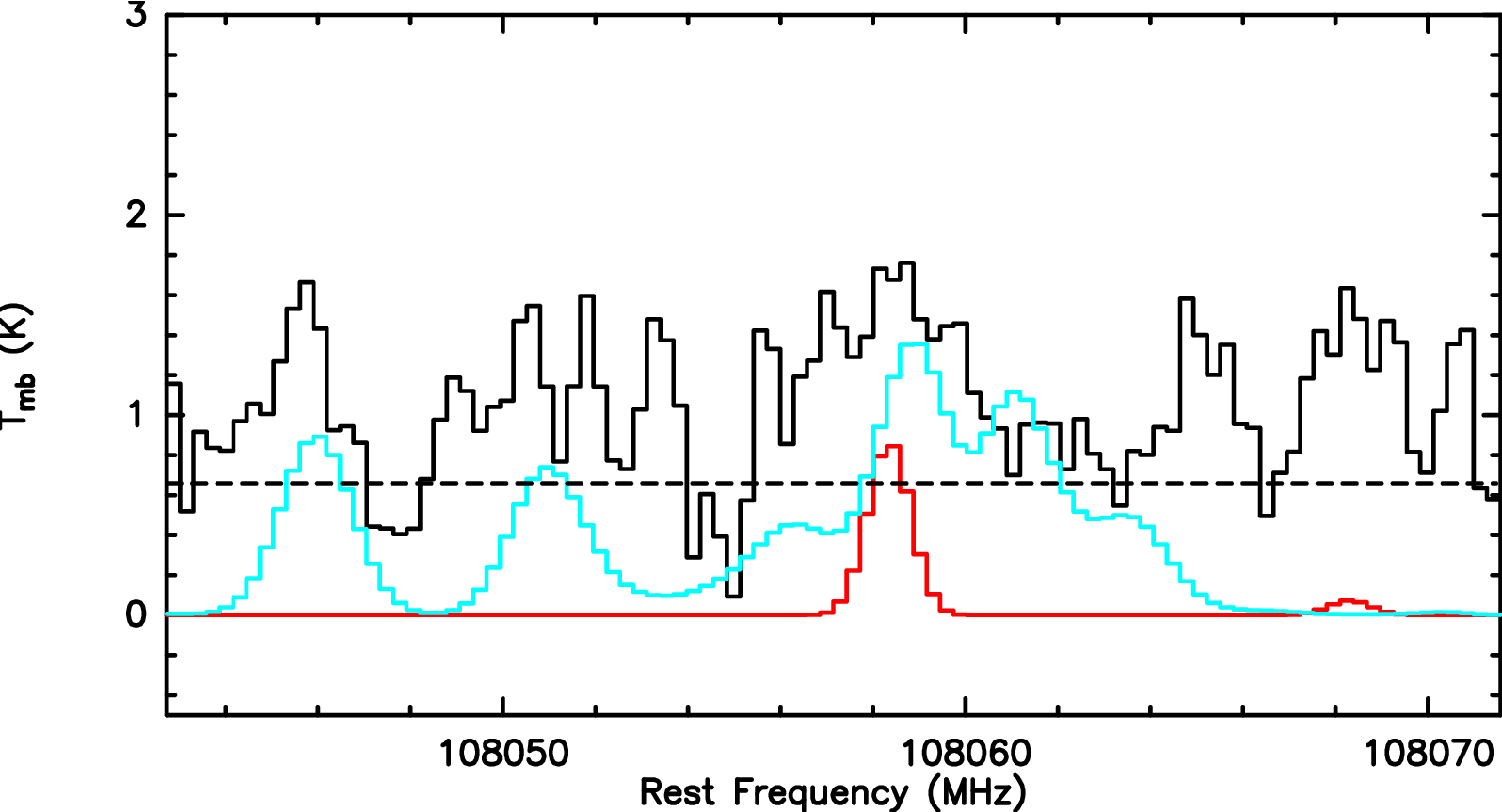} 
\includegraphics[width=2.5in]{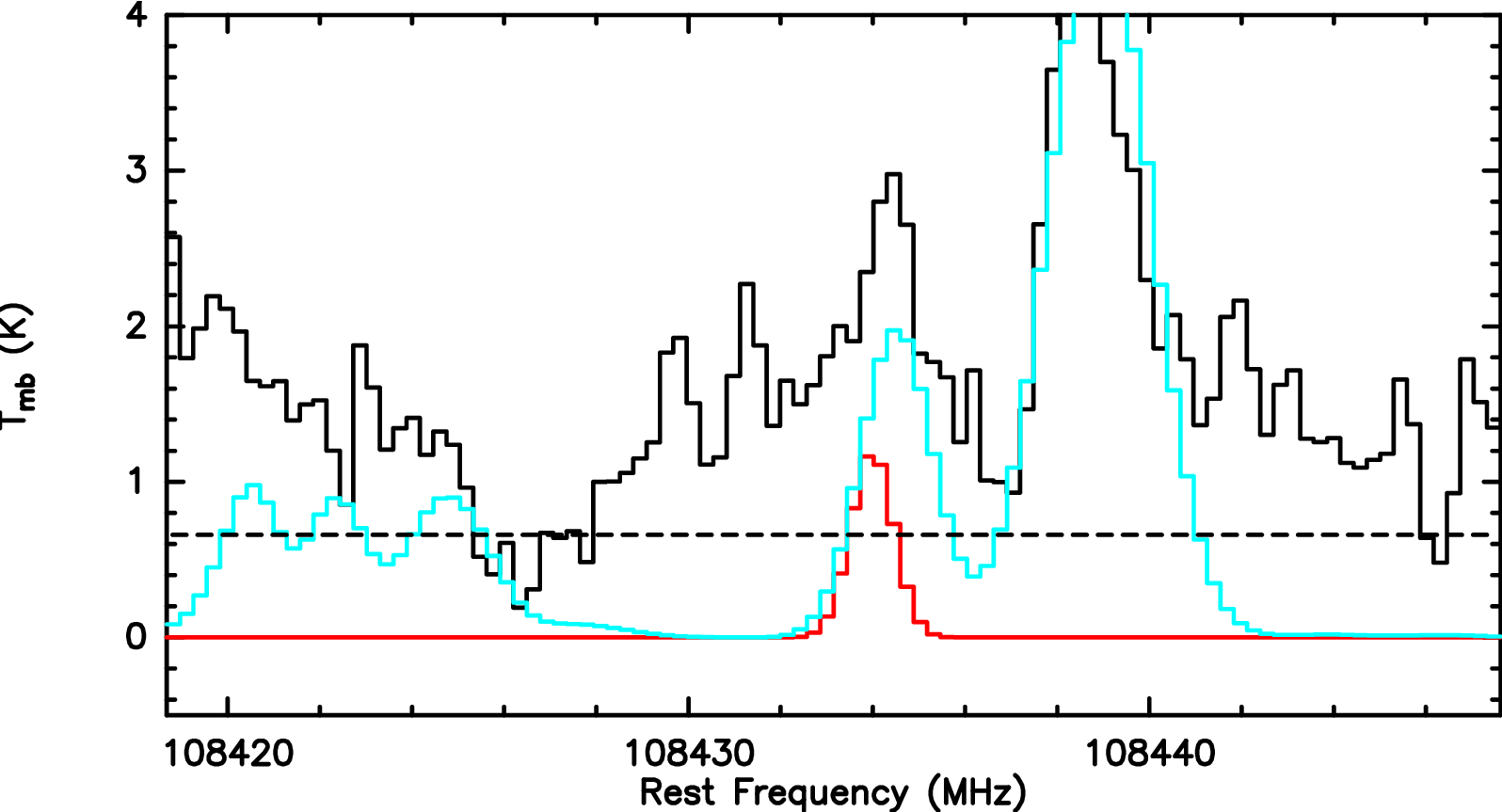}
 \includegraphics[width=2.5in]{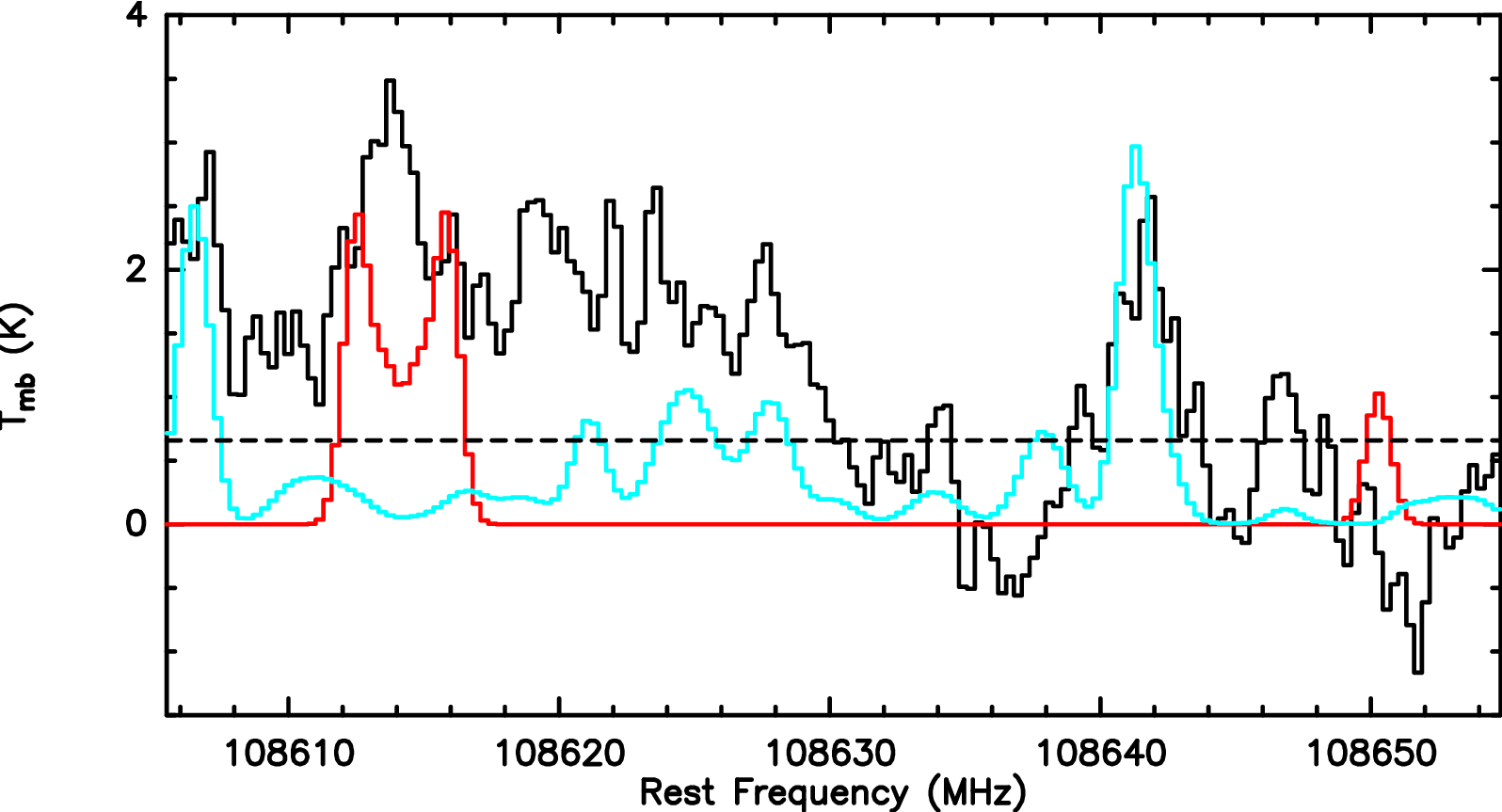}
 \includegraphics[width=2.5in]{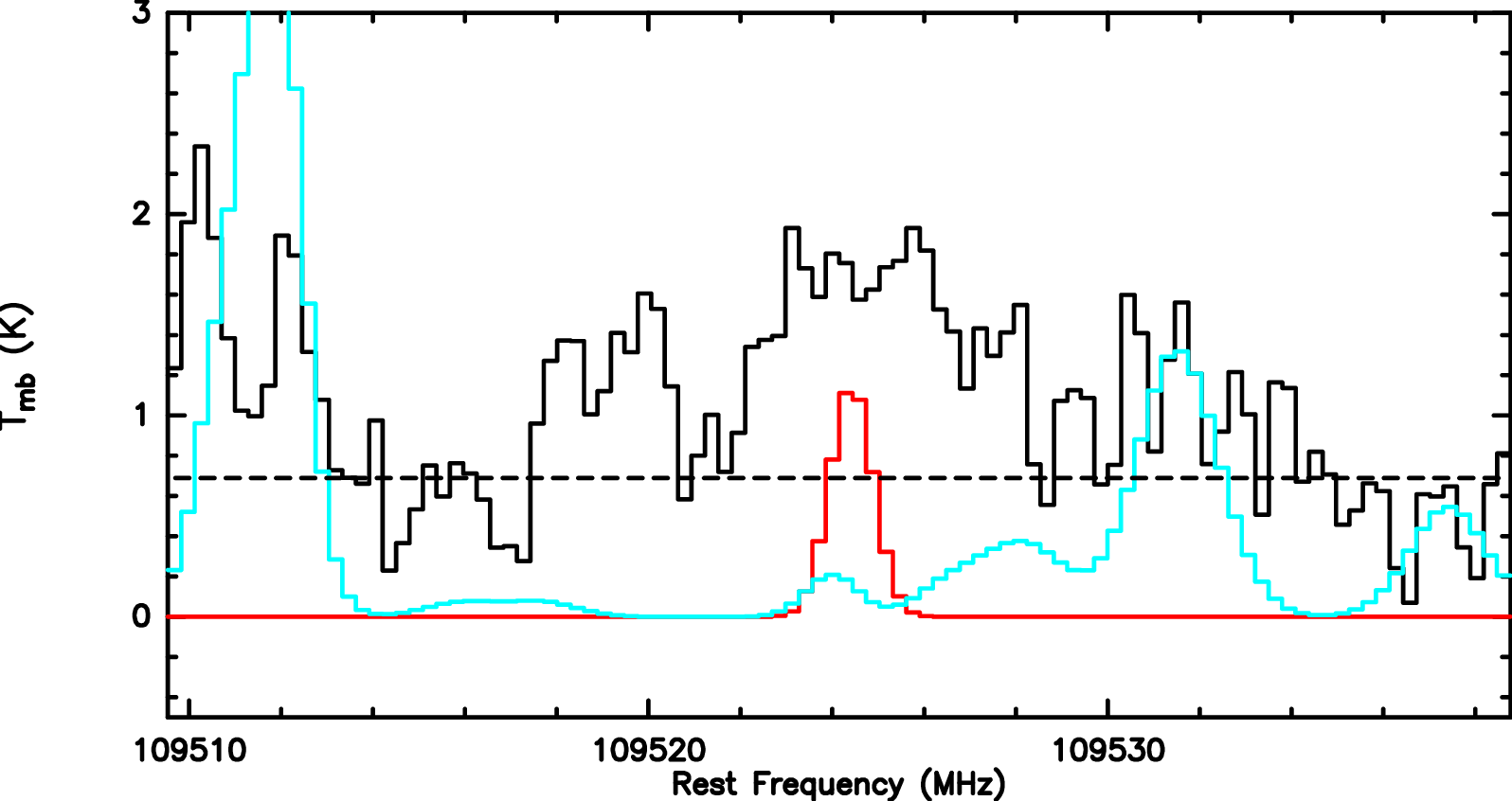}
  \includegraphics[width=2.5in]{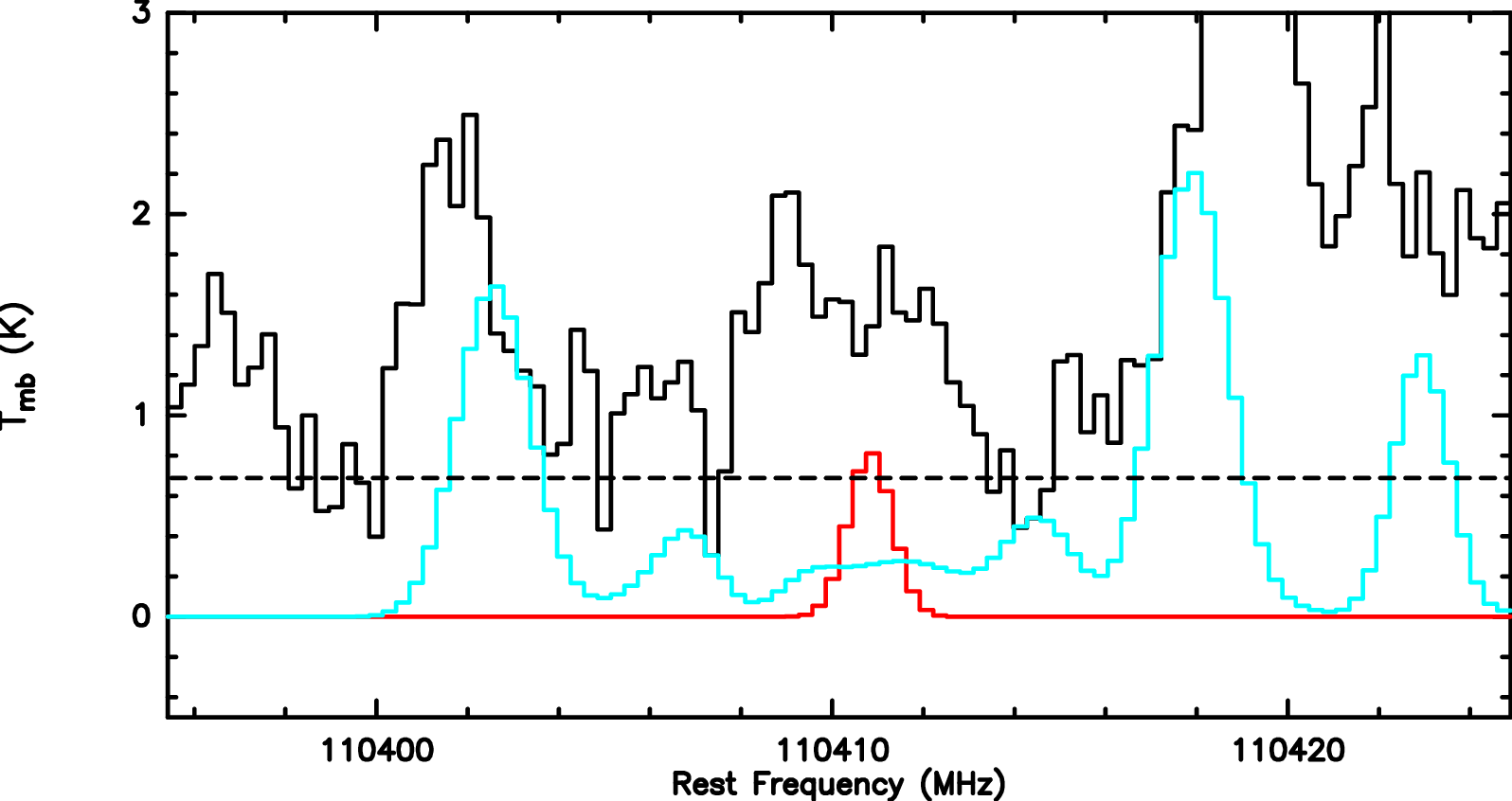}
}
\caption{Continued. Note that the NCCONH$_2$ transition at 108650 MHz is affected by $^{13}$CN absorption. }
\label{f:partial}
\end{figure}



\clearpage

\begin{figure}
\centering
{
 \includegraphics[width=2.5in]{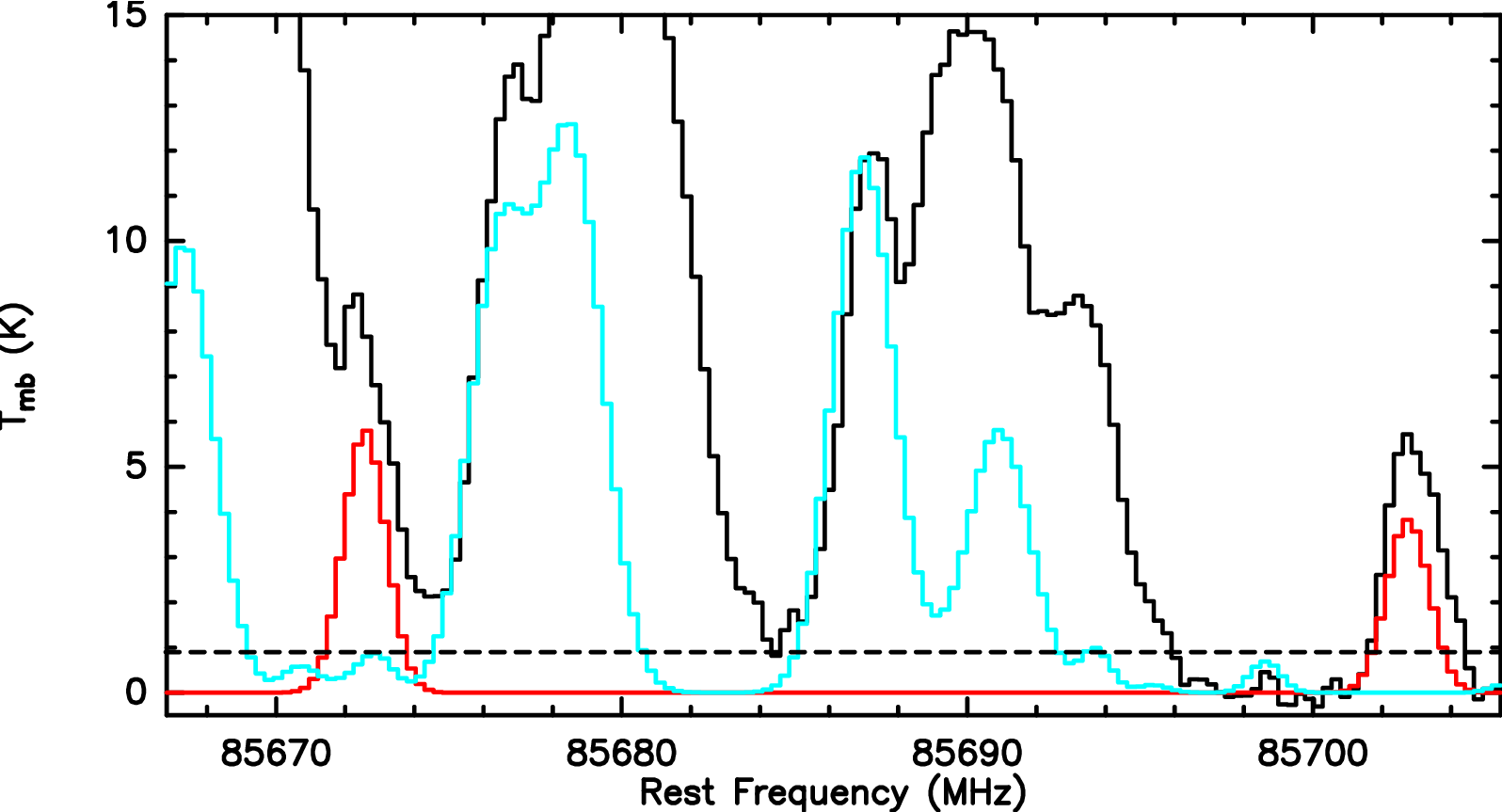}
  \includegraphics[width=2.5in]{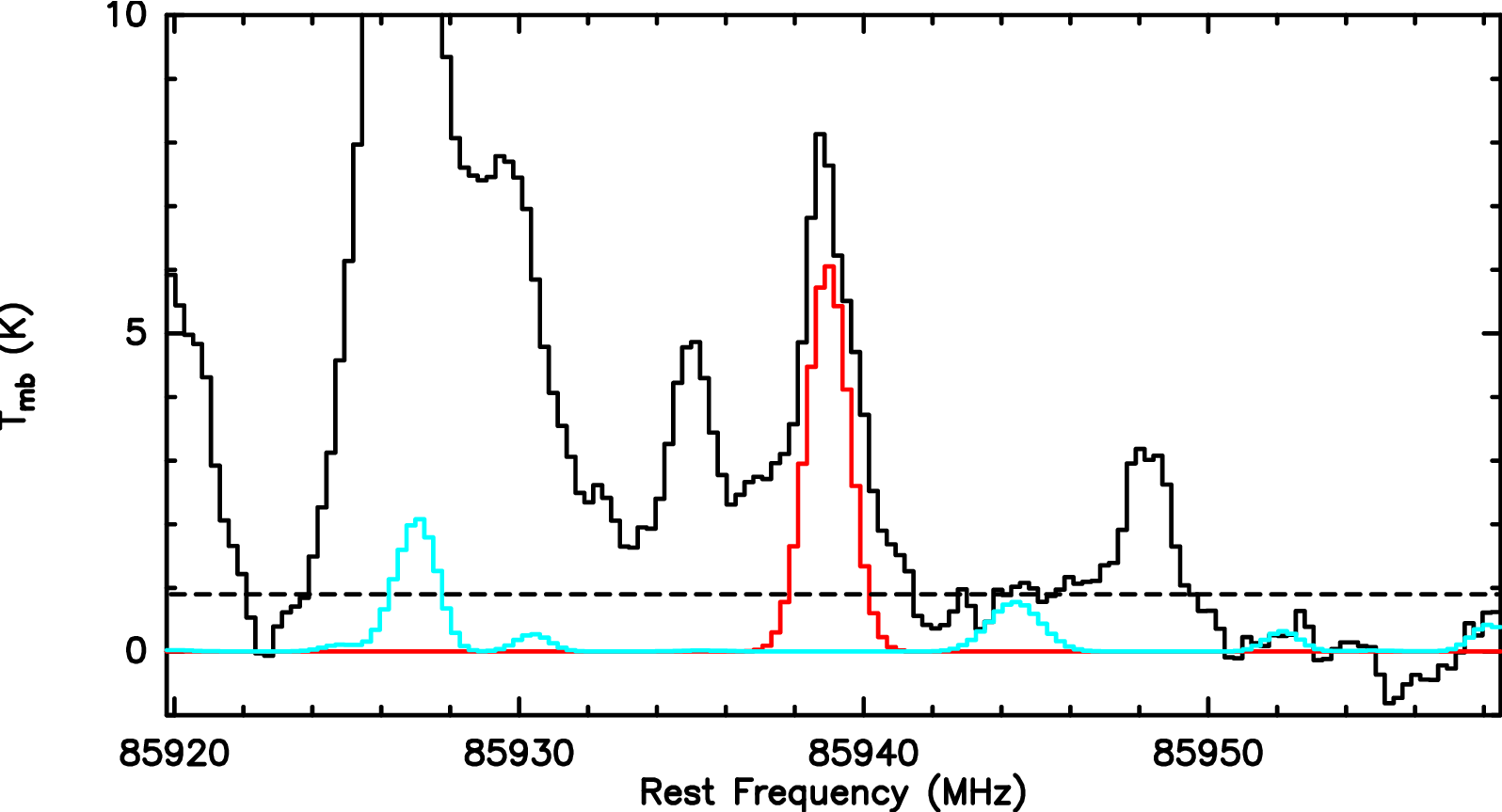}
  \includegraphics[width=2.5in]{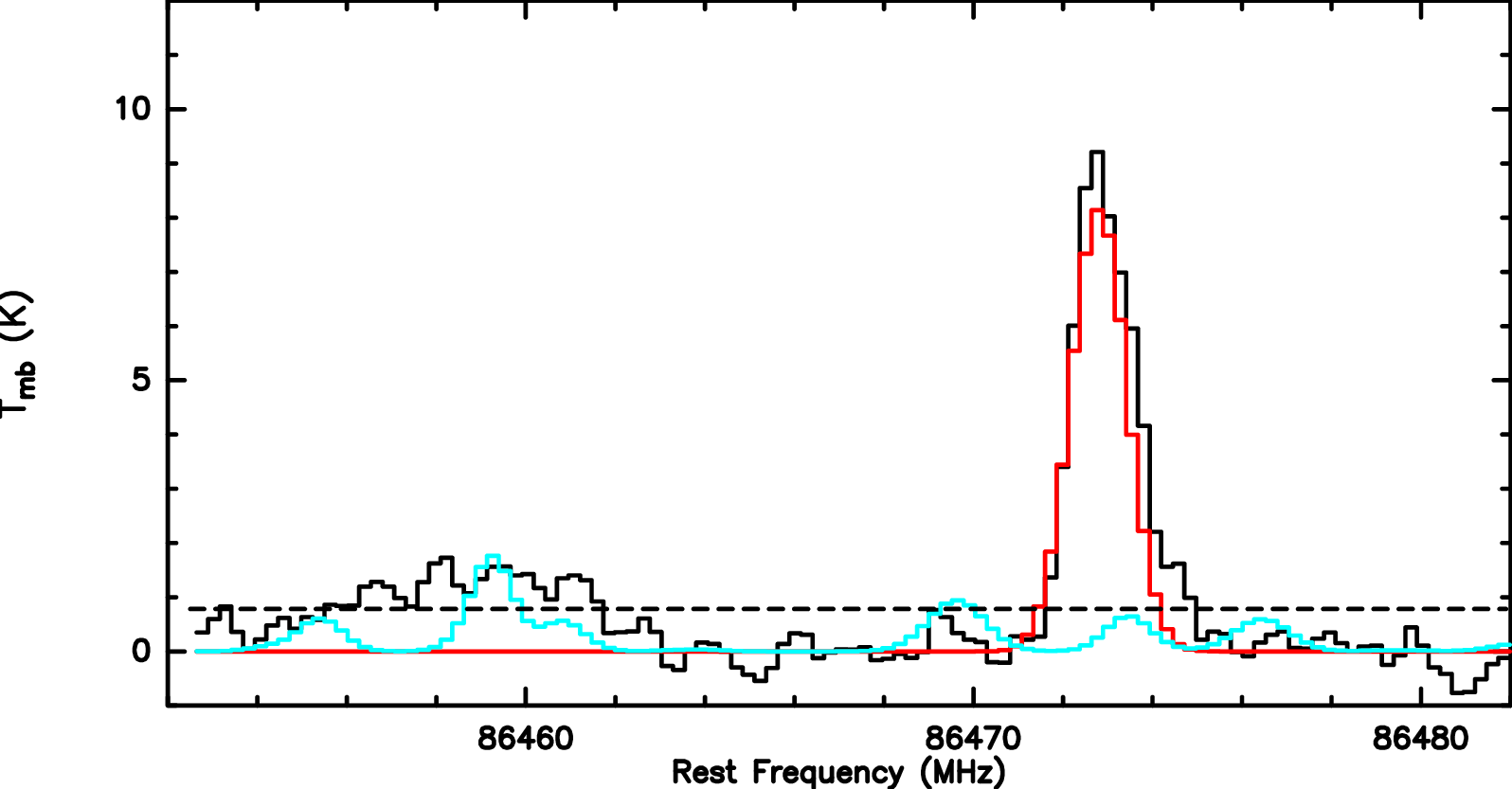}
   \includegraphics[width=2.5in]{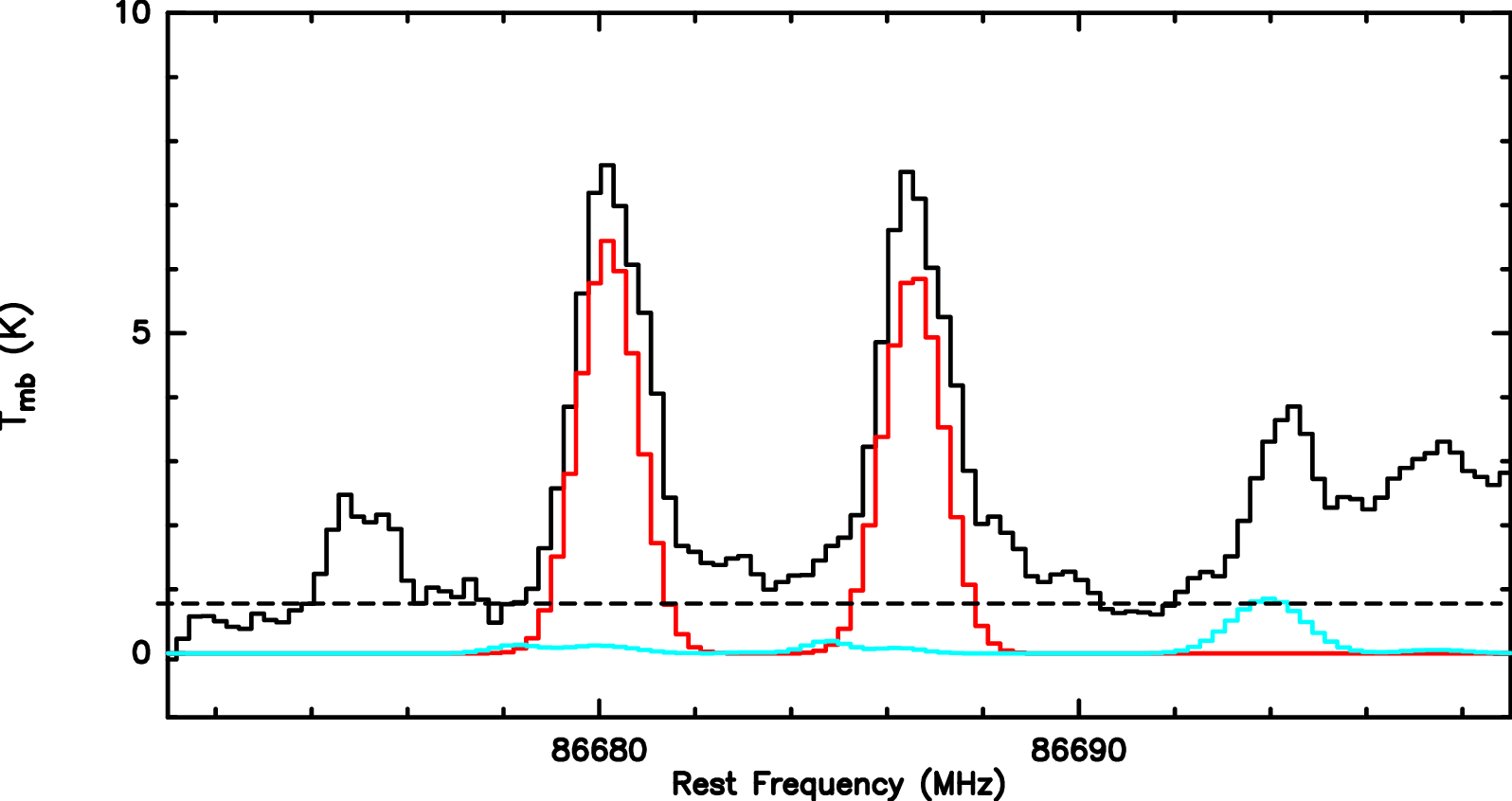}
    \includegraphics[width=2.5in]{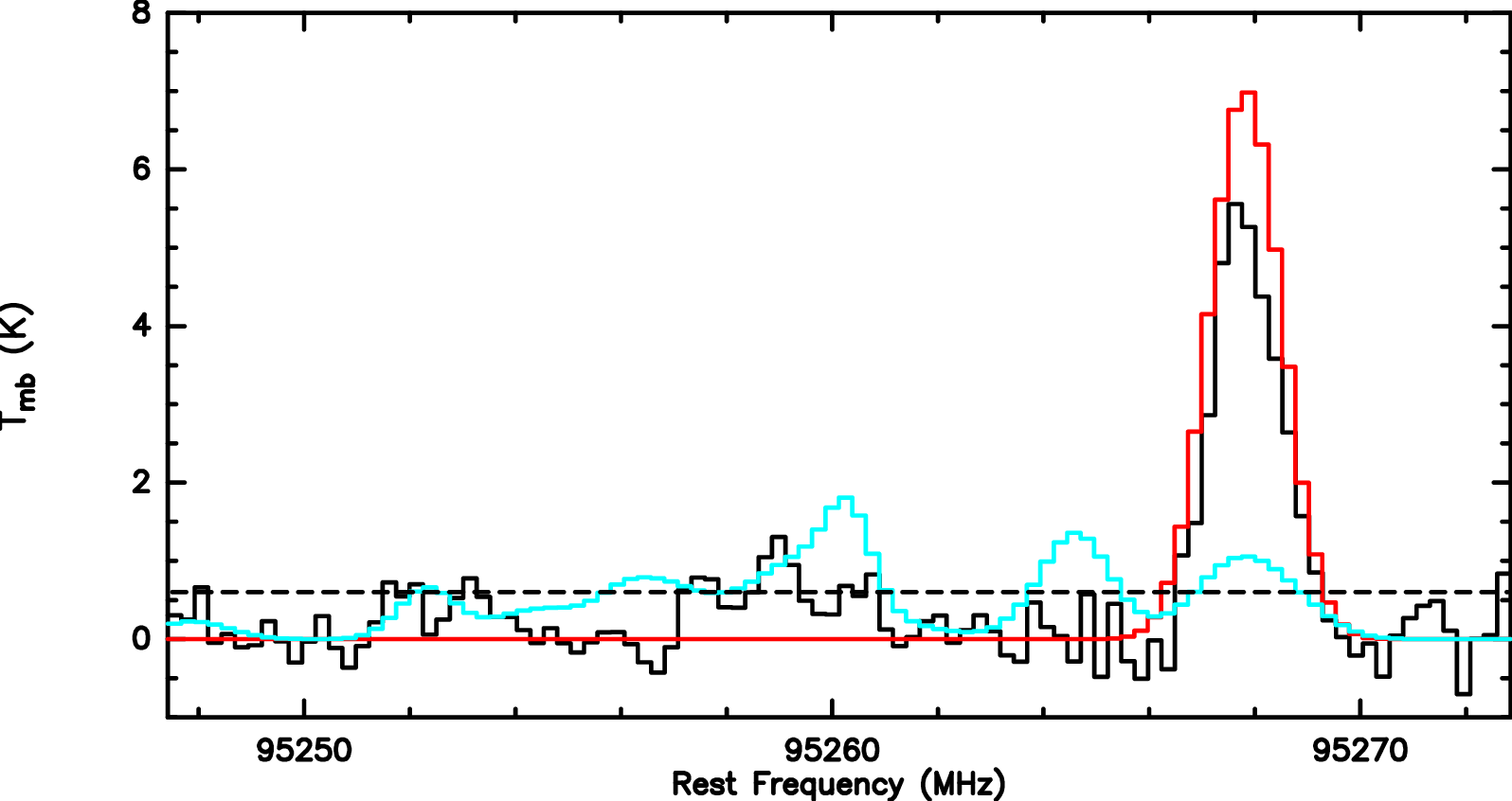}
     \includegraphics[width=2.5in]{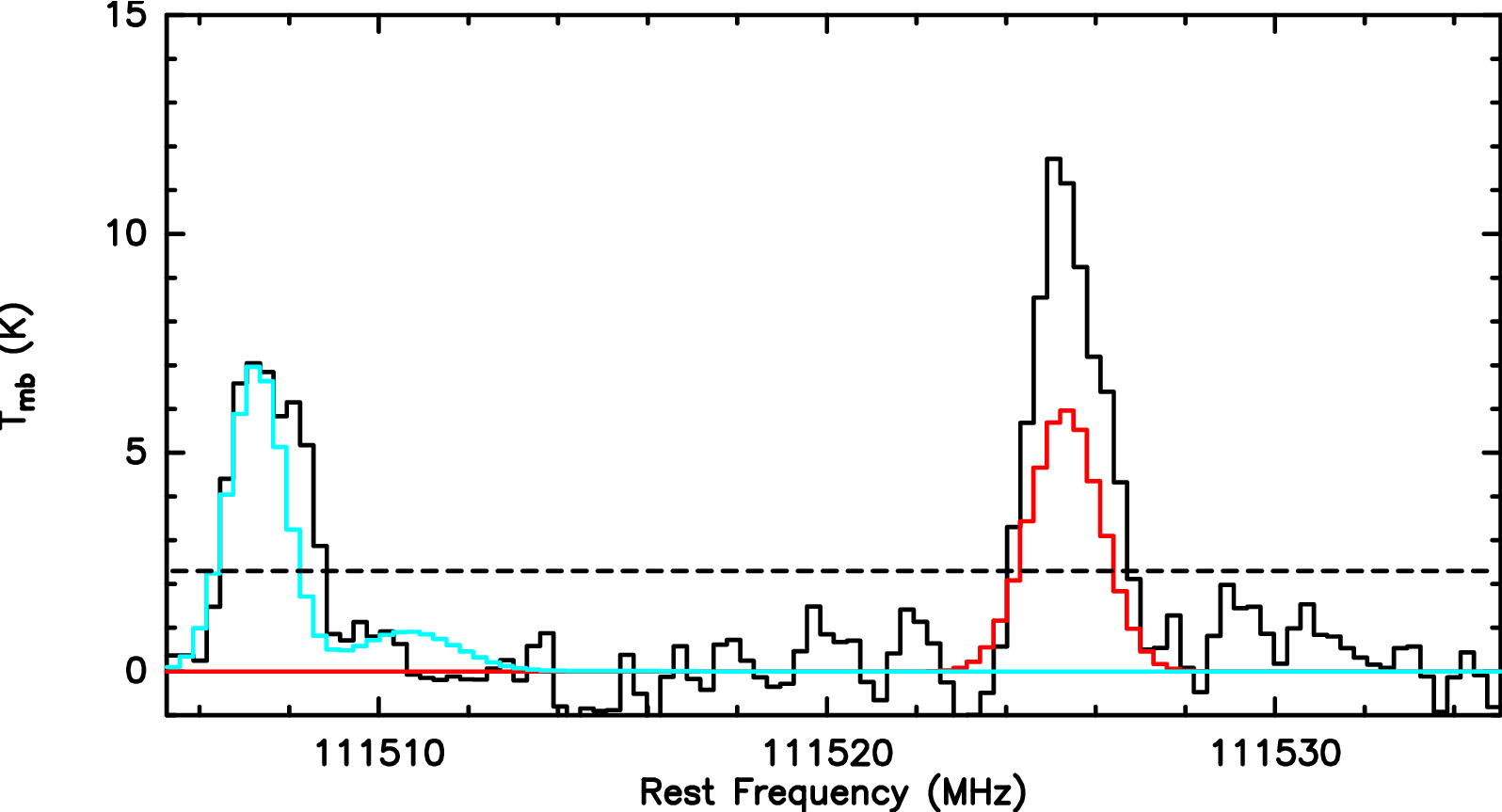}  
     \includegraphics[width=2.5in]{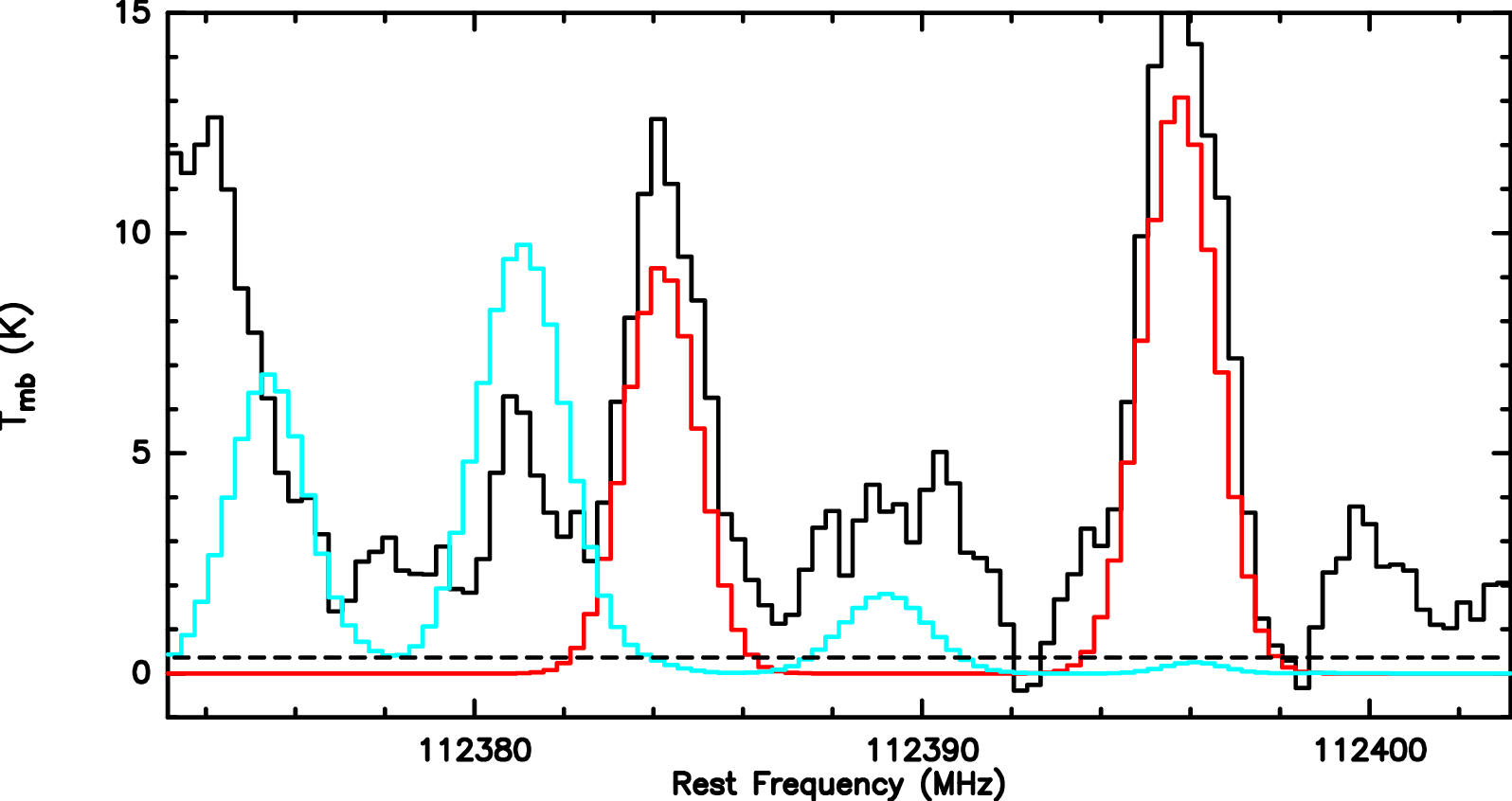}        
}
\caption{Transitions of CH$_3$NCO toward Sgr~B2(N1E). Black lines show spectrum observed with the ALMA telescope, while the red lines show the modeling results. The black dashed lines show the 3$\sigma$ noise levels. The cyan lines show the modeling results of other molecules.}
\label{f:ch3nco}
\end{figure}

\clearpage

\begin{figure}
\centering
{
 \includegraphics[width=2.5in]{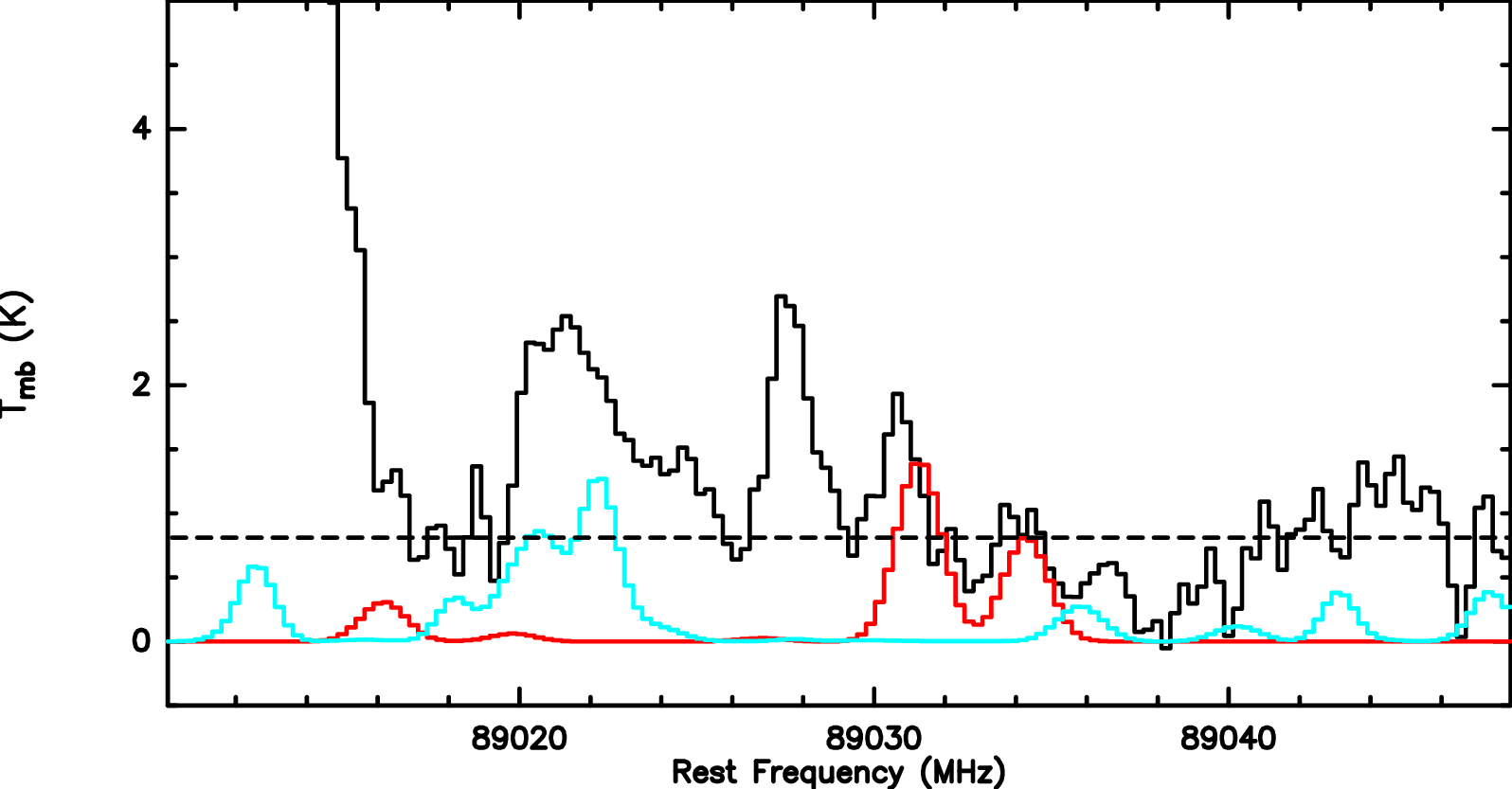}
  \includegraphics[width=2.5in]{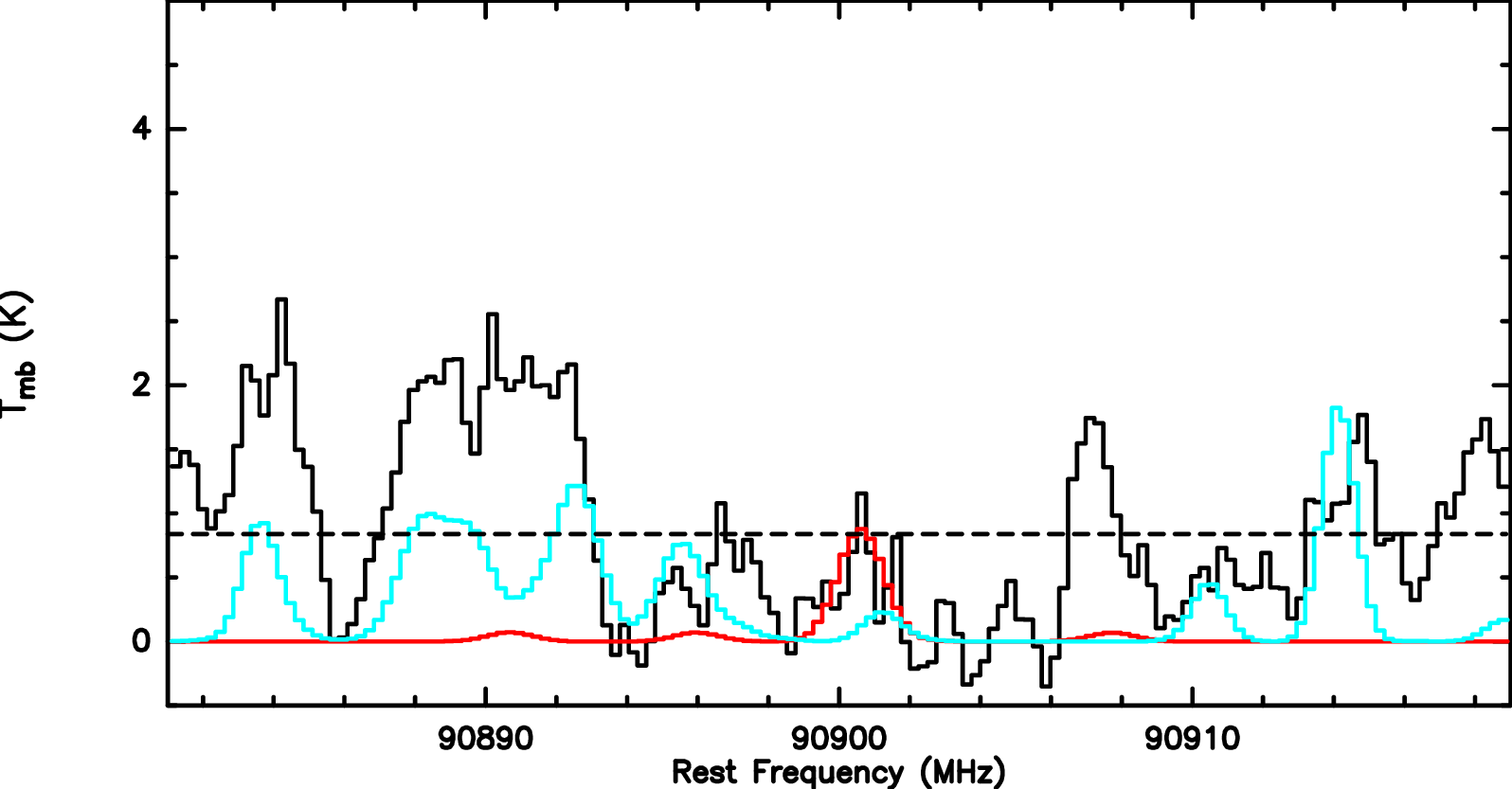}
 \includegraphics[width=2.5in]{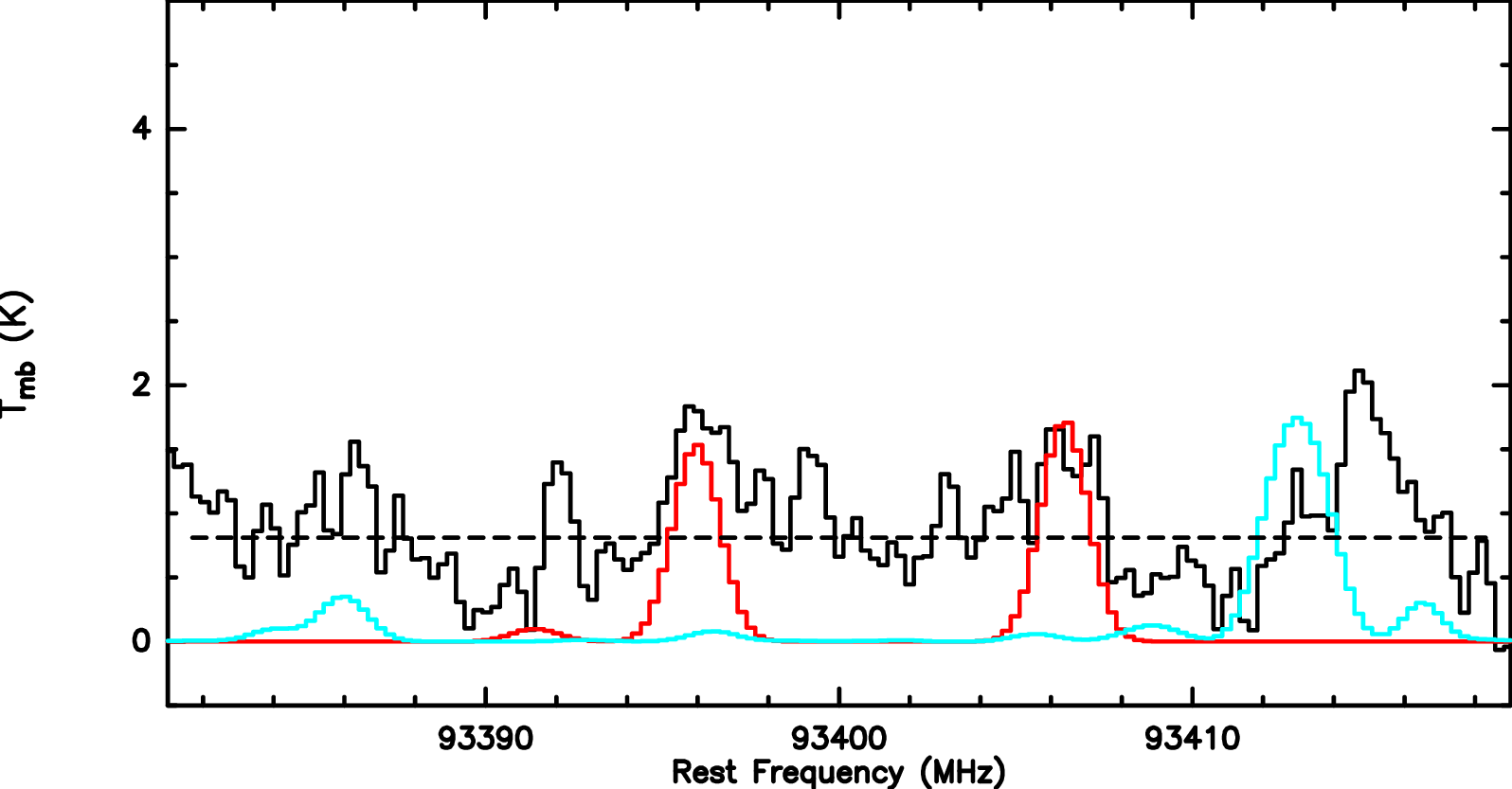}
  \includegraphics[width=2.5in]{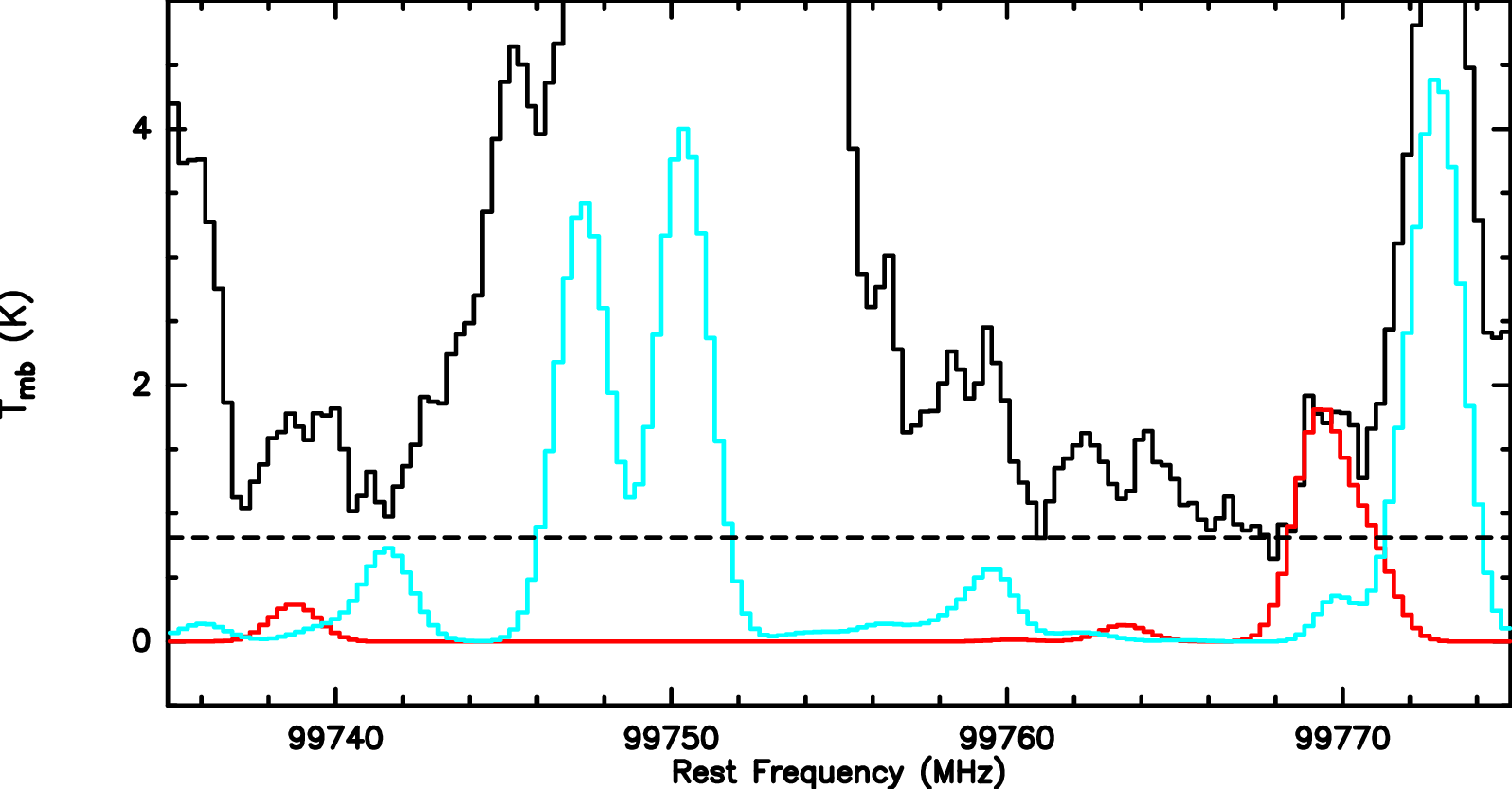}
  \includegraphics[width=2.5in]{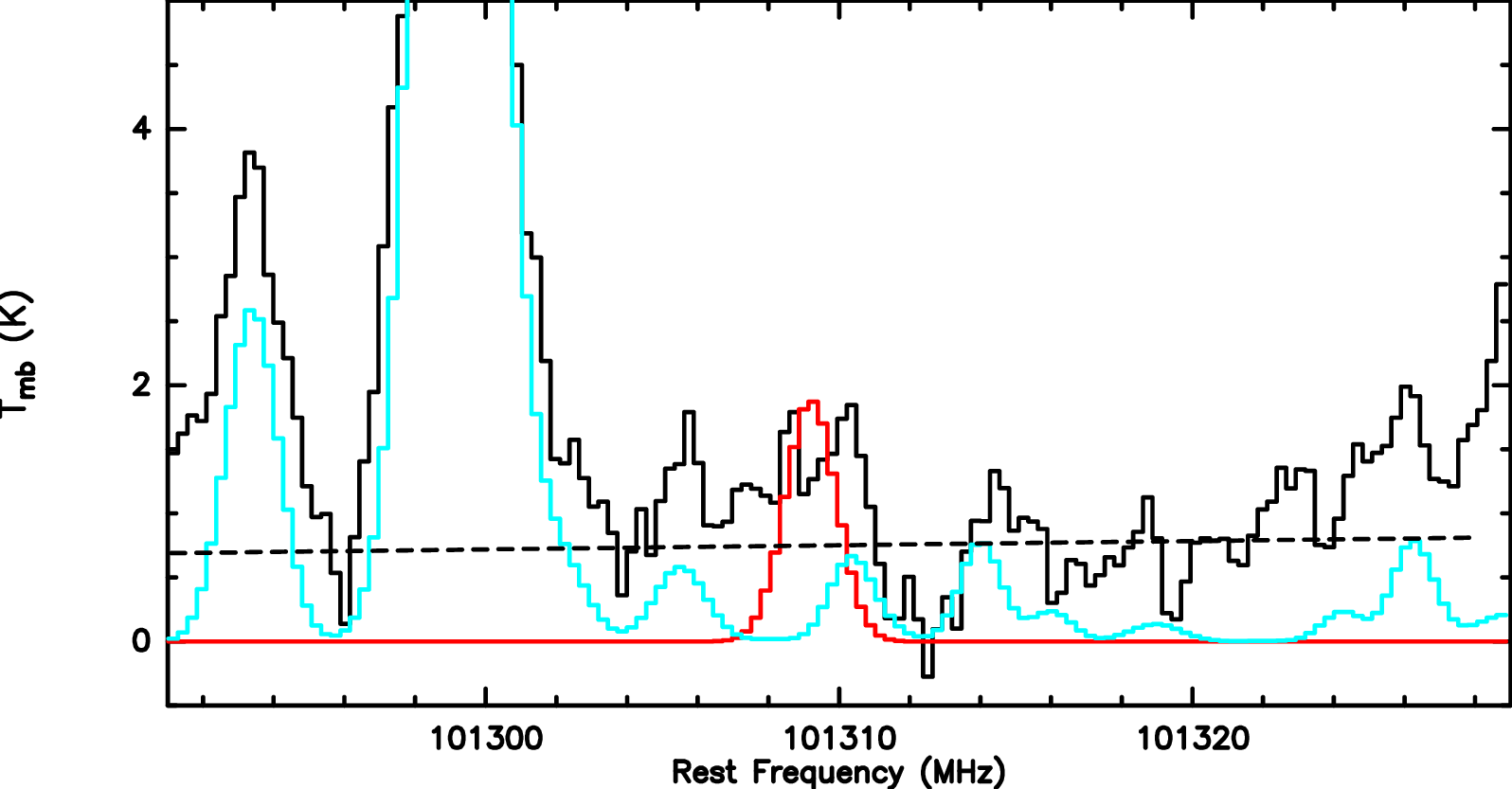}
    \includegraphics[width=2.5in]{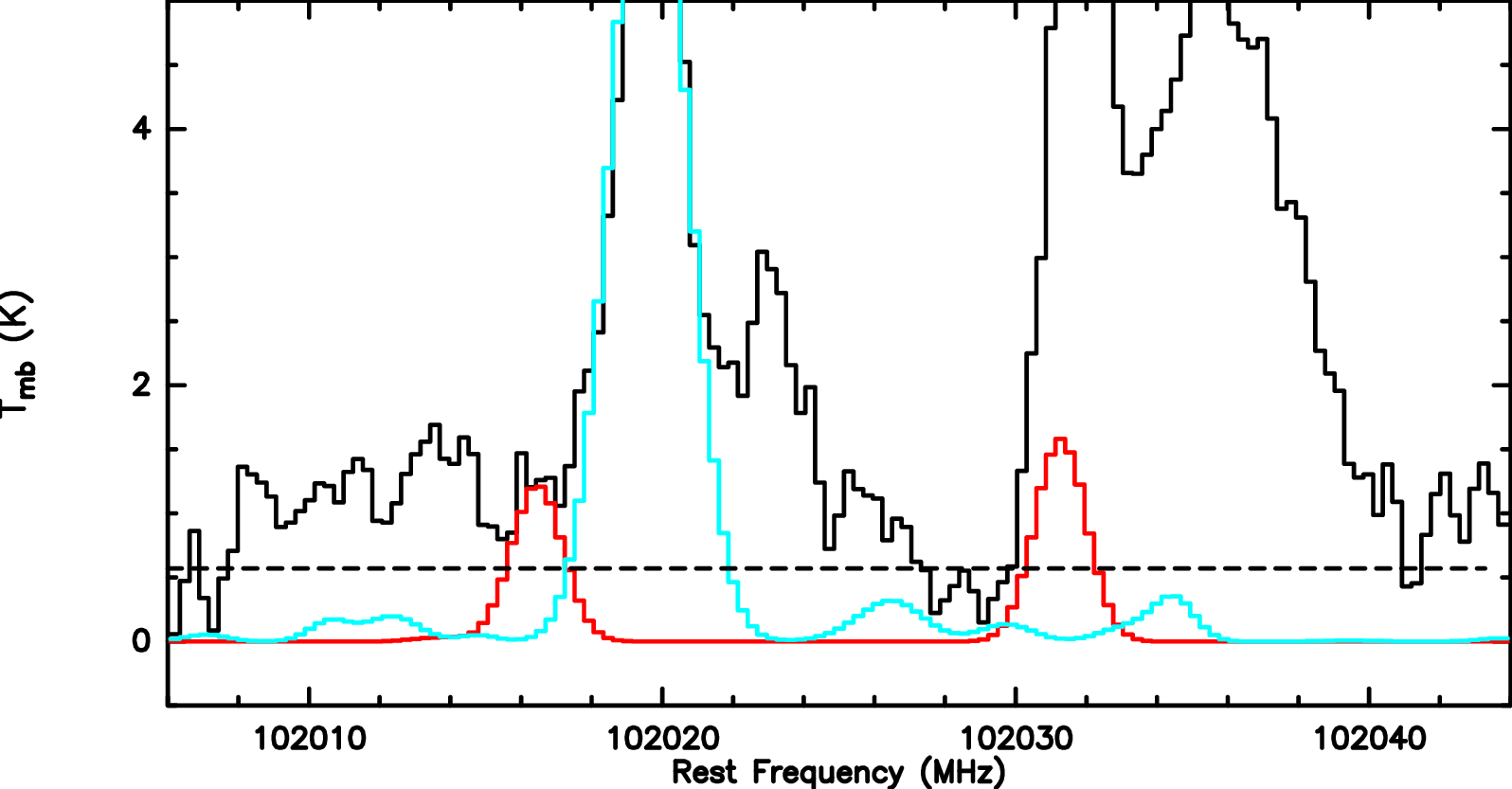}
   \includegraphics[width=2.5in]{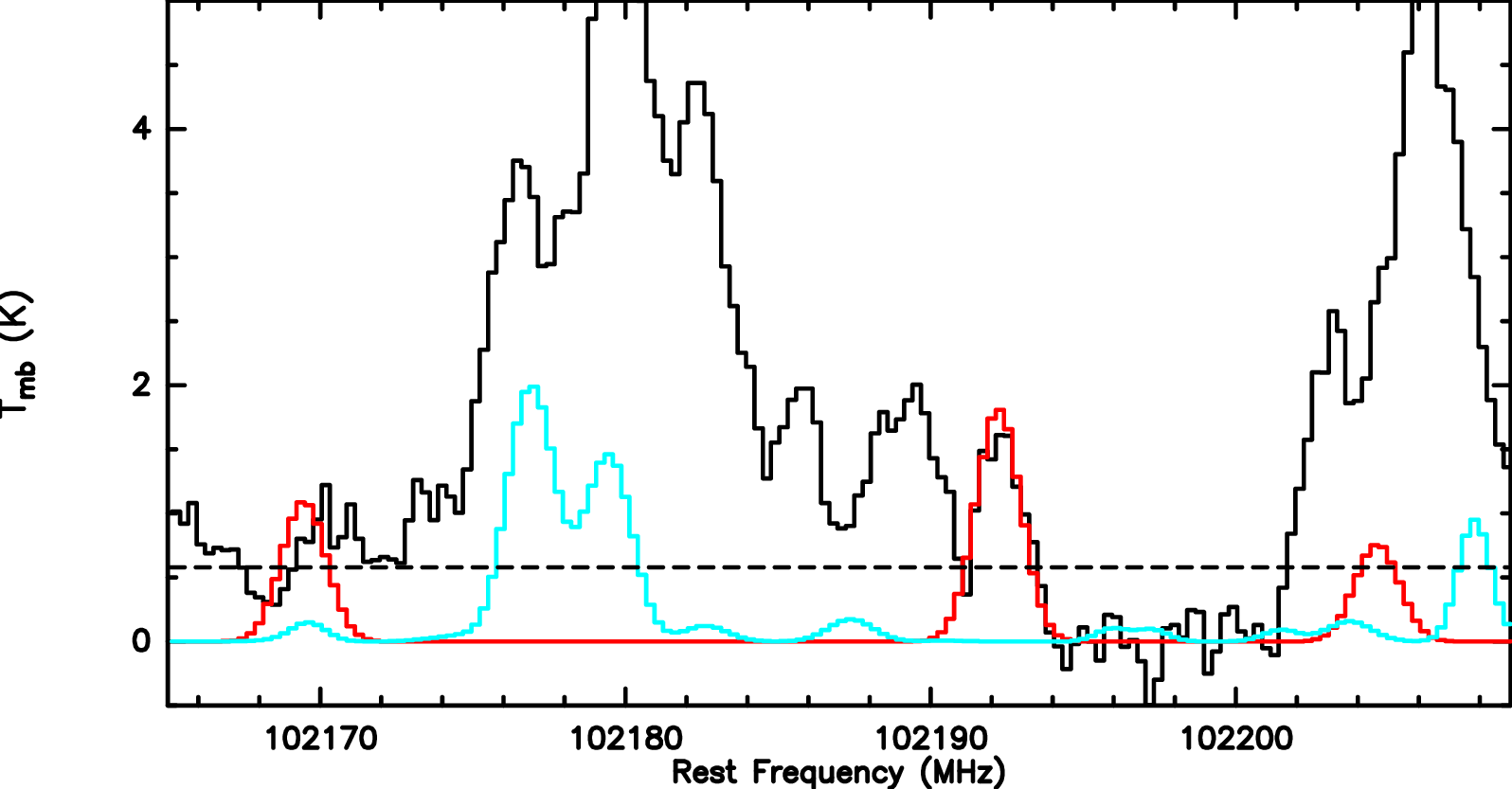}
    \includegraphics[width=2.5in]{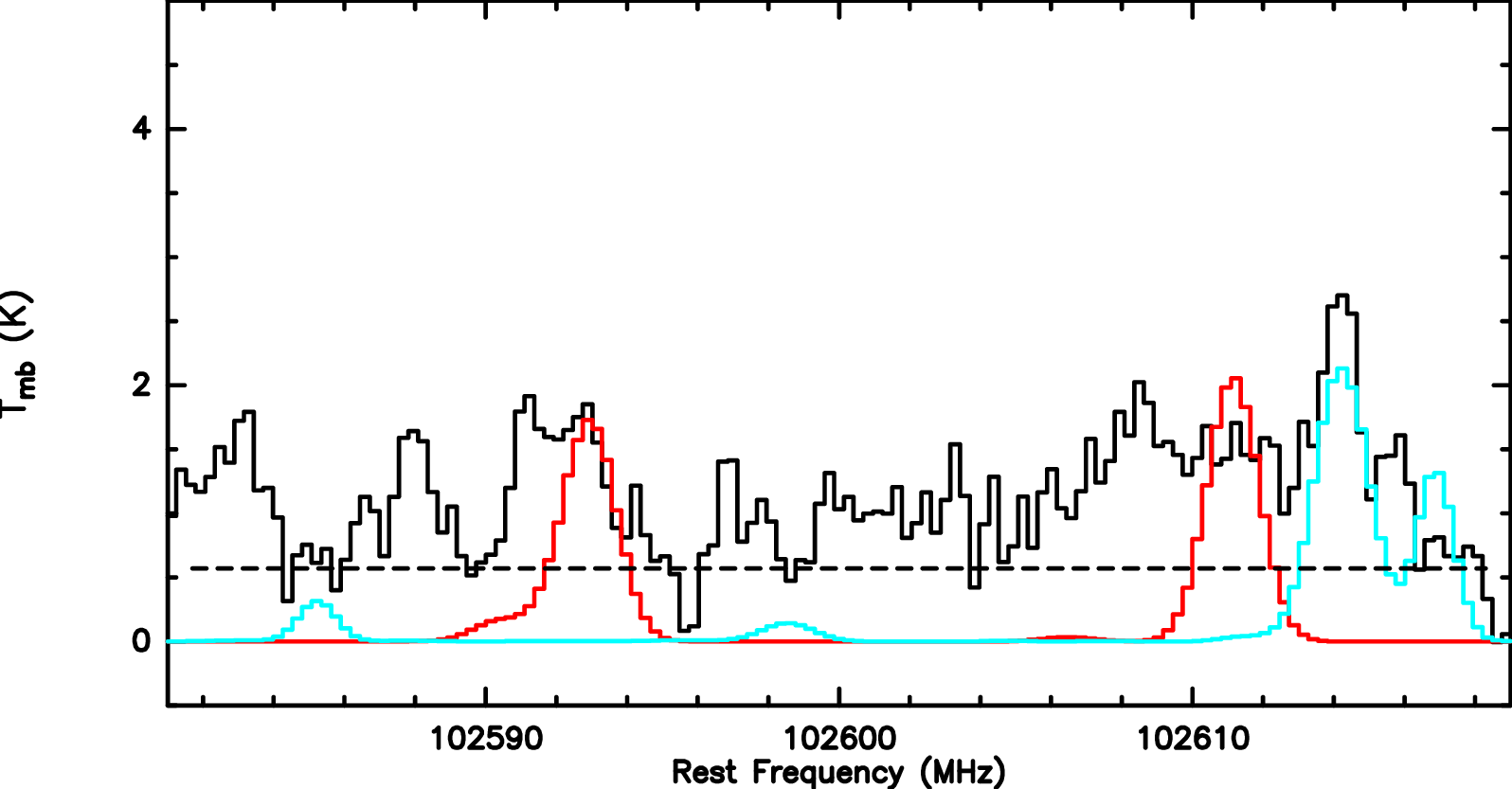} 
    \includegraphics[width=2.5in]{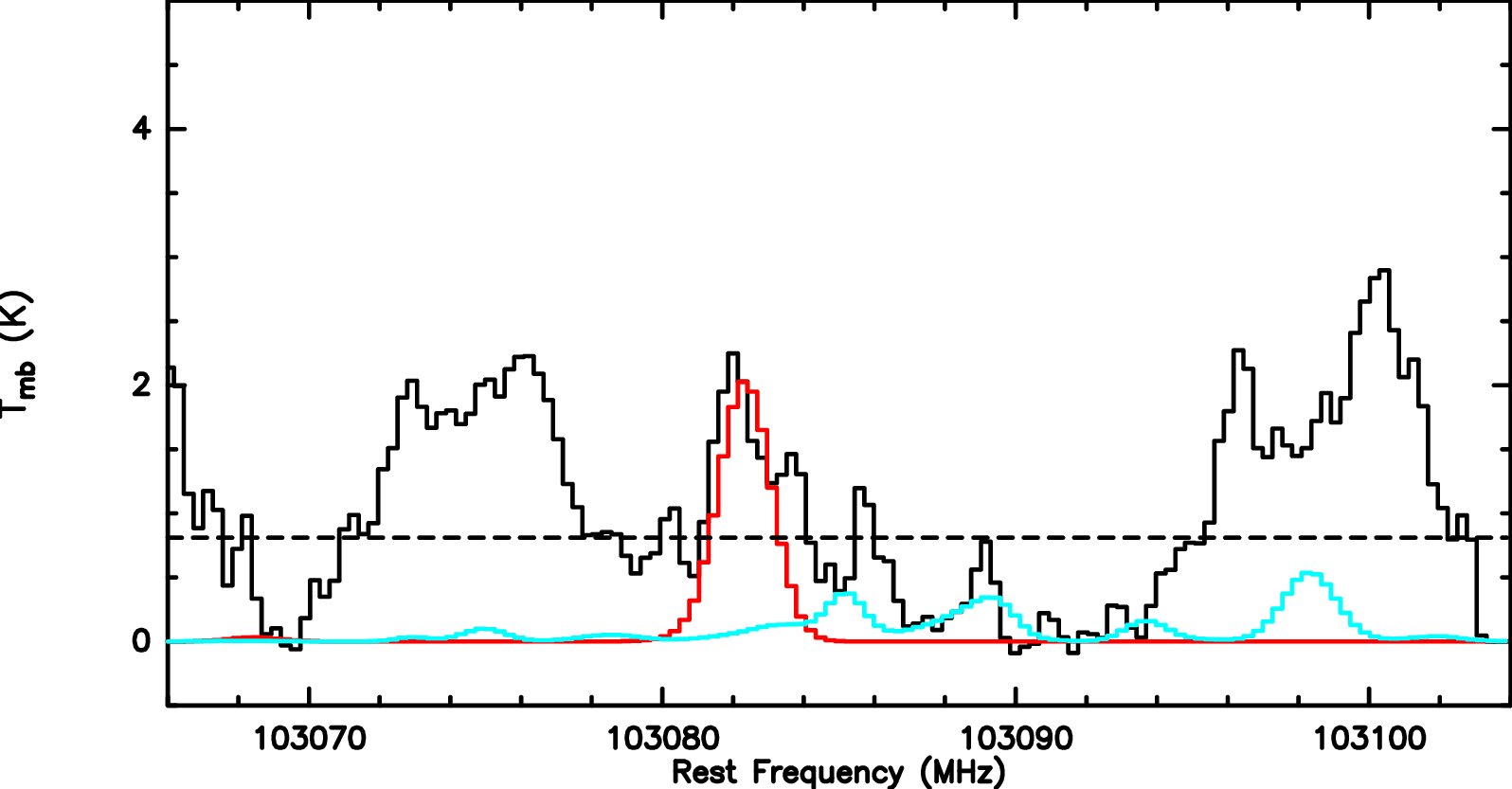}
}
\caption{Transitions of CH$_3$NHCHO toward Sgr~B2(N1E). The black solid lines show spectrum observed with the ALMA telescope, while the red lines show the modeling results. The black dashed lines show the 3$\sigma$ noise levels. The cyan lines show the modeling results of other molecules.}
\label{f:ch3nhcho}
\end{figure}

\clearpage

\begin{figure}
\centering
{\includegraphics[width=7.0in]{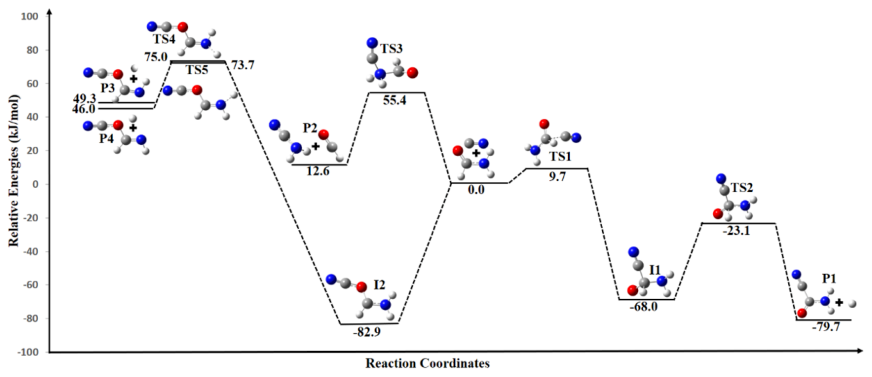}
}
\vspace{-285pt}
\caption{Potential energy surface for CN + NH$_2$CHO determined at the M06-2X/aug-cc-pVTZ//M06-2X/6-311+G(d,p) level of theory.
}
\label{f:energy}
\end{figure}

\clearpage

\begin{table}
\scriptsize
    \begin{center}
      \caption{\label{tab:freq} Rest frequencies and transition parameters of NCCONH$_2$ lines tentatively detected toward Sgr B2(N1E).}\label{tab:line}
      \begin{tabular}{lccccccc}
      \\
    \hline
    \hline
    
Number & Rest Freq.   &   Transition                                 & $E_{u}$  &  $\log_{10}A_{ul}$  &  $S_{ij} \mu^2$                \\
 &(MHz)         &   $J'(Ka', Kc') - J''(Ka'', Kc'')$    &   (K)        & (s$^{-1}$)                 &       (D$^2$)                              \\
\hline 
      &                                             &         unblended lines                       &                 &               &                     \\
(1)   &  86036.25670(0.0053) &  20(5, 16) - 20(4, 17)  &  85.0    &  -4.665    & 119.7           \\ 
(2)   &  94117.1956(0.0059)   &  19(7, 13)- 19(6, 14)   &  86.3   &  -4.505  & 125.6              \\ 
(3)    & 95286.1509(.0053)        &   23(7, 17)-23(6, 18)       & 117.7   &  -4.4649  &   160.0                 \\
(4)   &  95565.0503(.0061)         &  20(4,17) - 20(3, 18)   &  80.9   &   -4.645  &  91.5              \\ 
(5)   &  97481.6183(.0068)         &  13(7, 6) - 13(6, 7)   &  50.4   &  -4.553  &  70.027             \\          
(6)     &   99819.5119(0.0082)        &    	35(8, 27)-35(7, 28)    &     256.7    &   -4.359    &     268.3                \\       
(7)   &  106434.0041(0.0058)        &  24(8,17) - 24(7, 18)   &  131.6   &  -4.326  &  164.9              \\  
(8)   &   107207.9828(0.0067)       &    22(3, 19)-22( 2, 20)     &  95.7      &  -4.534     &    91.8         \\        
(9)   &  108243.4630(0.0056)        &  15(3, 13) - 14(3, 12)   &  45.7   &  -4.465  &  71.9            \\  
(10)   &  108734.3241(0.0057)        &  16(1,15) - 15(1, 14)   &  47.3   &  -4.449  &  78.4               \\               
\hline
       &                                             &     partially blended  lines                                   &                 &               &                     \\
 (1)     &  93247.9712(0.0064)         &     	20(3, 17)-20(2, 18)   &  80.7      &   -4.675            &    91.8                 \\       
(2)     &    93805.0949(0.0057)             &    20(7, 14)-20(6,15)       &     93.5   &  -4.500             &     134.9                \\
(3)       &   94272.7316(0.0054)        &    22(7, 16)-22(6, 17)      &   109.3    &    -4.481   &     152.4       \\
 (4)     &  94636.0867(0.0061)        &    18(7, 12)-18(6, 13)     &    79.4    &   -4.509     &        116.2     \\       
(5)       &   100344.5329(0.0066)     &    21( 3, 18)-21(2,19)     &    88.1    &  -4.600   &     91.8        \\       
(6)     &   101226.7711(0.0082)       &   32(7, 25)-32(6, 26)        &   214.0      &    -4.375           &    226.9       \\
(7)       &  102576.1442(0.0056)      &   15(1,14)-14(1,13)       &    42.1      &   -4.527     &       73.4      \\
(8)      &  103170.9109(0.0053)        &   13(3,10)-12(3, 9)       &    37.2    &    -4.529    &     62.5                \\     
(9)   &  103625.1683(0.0058)         &  25(6, 20) - 25(5, 21)   &  130.9   &  -4.414   &  151.9            \\     
(10)     &   105396.9363(0.0055)       &  15(2,13)-14(3,12)          &   45.6     &  -4.407       &  89.2                \\         
 (11)     &    105867.1182(0.0054)     &      14(5,10)-13(5, 9)       &  46.9      &   -4.531    &    61.9        \\       
 (12)    &    106600.4540(0.0057)        &      26(8,19)-26(7, 20)     &   150.3    &   -4.313     &  182.9          \\
 (13)      &  107618.0350(0.006)          &      22(8,15)-22(7,16)    &     114.4  &   -4.328   &     145.6       \\
(14)      &   108057.9882(0.0066)       &   22(4,19)-22(3, 20)         &    95.8     &    -4.524   &     91.7          \\       
(15)     &   108433.6177(0.0061)          &   	21(8,14)-21(7,15)        &  106.4     &    -4.329   &      136.0          \\          
(16)      &   108612.1374(0.0058)        &      17(0,17)-16(1,16)       &  48.1    &    -4.106  &   183.8         \\       
(17)     &     108613.2479(0.0058)       &       17(1,17)-16(1,16)      &   48.1   &   -4.440      &    85.1                 \\
  (18)     &   108614.3824(0.0058)       &     17(0,17)-16(0,16)    &     48.1      &    -4.440       &     85.1                \\
 (19)     &    108615.4928(0.0058)      &      17(1,17)-16(0,16)       &   48.1      &    -4.106  &    183.9        \\     
  (20)   &    109524.0529(0.0065)     &    19(8, 11)-19(7,12)        &    91.5      &     -4.338   &    117.0         \\       
(21)     &    110410.4446(0.0085)       &   36(8, 28)-36(7, 29)          &   270.4    &   -4.252      &     260.6                \\
  \hline              
      \end{tabular}
  \end{center}
  Notes. Col. (2): Rest frequencies with calculated uncertainties given in parentheses; Col. (3): Transition; Col. (4): Upper level energy; Col. (5): Base 10 logarithm of the Einstein $A_{ul}$ coefficient; Col. (6): Line strength $\mathrm{S}_{ij}\mu^2$.
\end{table}

\clearpage

\begin{table}
\scriptsize
    \begin{center}
      \caption{Parameters of LTE model of NCCONH$_2$ and related molecules toward Sgr B2(N1E).}\label{tab:column}
      \begin{tabular}{lccccccc}
      \\
    \hline   
    \hline
    
Molecule  &  Size     &     $T_{rot}$   &  $N_{N1E}$   &    $\Delta V$         &       $V_{off}$                     &   $N_{N1S}^{**}$     \\
                &   (")        &   (K)              &   ($cm^{-2}$)   &    ($km s^{-1}$)      &    ($km s^{-1}$)     &    ($cm^{-2}$)       \\
\hline              
NCCONH$_2$, v=0        &   2.3      &    150       &   2.4(15)            &    3       &       -1     &  -     \\         
CH$_3$NCO               &    2.3     &     150       &    2.4(16)             &   5      &     0         &     2.5(17)    \\
HCONH$_{2}$, v=0$^*$   &    2.3     &     150     &     2.8(17)         &    6       &      -1         &  2.9(18)                \\ 
CH$_3$CONH$_{2}$, v=0$^*$ &   2.3    &     150     &     2.5(16)    &    3.6       &      -1      &   4.1(17)     \\      
CH$_3$NHCHO, v=0    &   2.3   &    150   &   2.1(16)                     &     5       &       0           &  2.6(17)      \\    
\hline
      \end{tabular}
  \end{center}
  Notes. Col. (1): Molecule name; Col. (2): Source diameter (FWHM), Col. (3): Rotational temperature; Col. (4): Total column density of the molecule. x(y) means $x \times 10^y$; Col. (5): Linewidth (FWHM); Col. (6): Velocity offset with respect to the assumed systemic velocity of Sgr B2(N1), $V_{sys}=64$ km s$^{-1}$; Col (7): column density of the molecule toward Sgr B2(N1S) \citep{Belloche19}. $^*$: cited from \citet{Li21}. $^{**}$: cited from \citet{Belloche19}. 
\end{table}

{}

\end{document}